\documentclass[a4paper,11pt]{article}
\pdfoutput=1 
\usepackage{jheppub} 
\usepackage[T1]{fontenc} 
\usepackage{caption}
\usepackage{subcaption}

\RequirePackage[numbers,sort&compress]{natbib}
\usepackage{cleveref}
\Crefname{equation}{eq.}{eqs.}
\Crefname{section}{section}{sections}
\Crefname{figure}{figure}{figures}
\Crefname{appendix}{appendix}{appendices}

\usepackage{braket}
\usepackage{float}
\usepackage{amsmath}
\usepackage{amssymb}
\usepackage{mathtools}
\usepackage{siunitx}
\usepackage{amsfonts}
\usepackage{dsfont}
\usepackage{array}
\usepackage{bm}
\usepackage{mathrsfs}
\usepackage{pifont}
\usepackage{multirow}
\usepackage{upgreek}
\usepackage{verbatim}
\usepackage{slashed}
\usepackage{ucs}
\usepackage{afterpage}
\graphicspath{{./pics/}}

\newcommand{\nn}{\nonumber}

\newcommand{\cM}[0]{\mathcal M}

\newcommand{\cK}[0]{\mathcal K}

\newcommand{\PV}[0]{{\rm PV}}
\newcommand{\iso}[0]{{\rm iso}}

\newcommand{\Kdf}[0]{{\cK_{3}}}  

\makeatletter
\renewcommand*\env@matrix[1][\arraystretch]{%
  \edef\arraystretch{#1}%
  \hskip -\arraycolsep
  \let\@ifnextchar\new@ifnextchar
  \array{*\c@MaxMatrixCols c}}
\makeatother

\title{\boldmath 
Electroweak three-body decays in the presence of two- and three-body bound states
}

\author[a,b]{Raul A. Brice\~no}
\author[c]{, Andrew W. Jackura}
\author[a,b]{, Dimitra A. Pefkou}
\author[d]{, and Fernando Romero-L\'opez}

\affiliation[a]{Department of Physics, University of California, Berkeley, CA 94720, USA}
\affiliation[b]{Nuclear Science Division, Lawrence Berkeley National Laboratory, Berkeley, CA 94720, USA}
\affiliation[c]{Department of Physics, 
William \& Mary, 
Williamsburg, VA 23187, USA}
\affiliation[d]{Center for Theoretical Physics, Massachusetts Institute of Technology, Cambridge, MA 02139, USA}

\emailAdd{rbriceno@berkeley.edu}
\emailAdd{awjackura@wm.edu}
\emailAdd{dpefkou@berkeley.edu}
\emailAdd{fernando@mit.edu}

\abstract{Recently, formalism has been derived for studying electroweak transition amplitudes for three-body systems both in infinite and finite volumes. The formalism provides exact relations that the infinite-volume amplitudes must satisfy, as well as a relationship between physical amplitudes and finite-volume matrix elements, which can be constrained from lattice QCD calculations. This formalism poses additional challenges when compared with the analogous well-studied two-body equivalent one, including the necessary step of solving integral equations of singular functions.  In this work, we provide some non-trivial analytical and numerical tests on the aforementioned formalism. In particular, we consider a case where the three-particle system can have three-body bound states as well as bound states in the two-body subsystem. For kinematics below the three-body threshold, we demonstrate that the scattering amplitudes satisfy unitarity. We also check that for these kinematics the finite-volume matrix elements are accurately described by the formalism for two-body systems up to exponentially suppressed corrections. Finally, we verify that in the case of the three-body bound state, the finite-volume matrix element is equal to the infinite-volume coupling of the bound state, up to exponentially suppressed errors. }

\preprint{MIT-CTP/5683}

\begin{document} 
\maketitle
\flushbottom

\section{Introduction}
\label{sec:intro}

Accurately computing electroweak transitions involving hadronic states is crucial for testing the Standard Model of Particle Physics. For example, the current tension between the theoretical prediction and experimentally measured values of the muon anomalous magnetic moment has driven the community to assess the largest theoretical uncertainties coming from hadronic intermediate states~\cite{Aoyama:2020ynm}. Electroweak transitions also serve to probe the non-perturbative quark and gluon interactions within Quantum Chromodynamics (QCD) itself, allowing one to determine the substructure of and understand the role of glue in matter, one of the central motivations behind the experimental program being planned for the upcoming Electron-Ion Collider~\cite{AbdulKhalek:2021gbh}.

While the electroweak sector is amenable to perturbation theory for computations of a given observable, processes involving strongly interacting hadrons requires systematically improvable, non-perturbative computational methods.
An approach to non-perturbative QCD known as lattice QCD allows one to numerically compute low-energy observables in finite, discrete spacetime volume. While the computational costs and theoretical support pose significant challenges, the growing number of results involving scattering and electroweak matrix elements have shown that quantitative results of low-energy strong dynamics are possible~\cite{Leskovec:2024pzb,Meinel:2024pip}.

An important aspect in reconstructing low-energy observables of few-hadron systems from lattice QCD is correcting for finite-volume effects inherent in any calculation. Observables involving multiple hadrons, such as electroweak matrix elements, suffer from power-law scaling effects when computed in a finite spatial volume~\cite{Luscher:1986pf, Luscher:1990ux,Lellouch:2000pv}. Isolating the power-law corrections to these observables allows one to construct exact non-perturbative mappings between finite-volume observables, such as energy spectra and matrix elements, to corresponding infinite-volume reaction amplitudes, effectively removing the volume artifacts.

The first work to study the effects for matrix elements was by  Lellouch and L\"uscher~\cite{Lellouch:2000pv}, who found an exact relationship between the $K\to \pi\pi$ decay amplitude and the corresponding finite-volume matrix elements of the weak current. This has since been generalized to increasingly complicated reactions involving two-particle states in the initial and/or final state~\cite{Lin:2001ek,Detmold:2004qn,Kim:2005gf,Christ:2005gi,Meyer:2011um,Hansen:2012tf,Briceno:2012yi,Bernard:2012bi,Agadjanov:2014kha,Briceno:2014uqa,Briceno:2015csa,Briceno:2015tza,Baroni:2018iau,Briceno:2019nns,Briceno:2020xxs,Feng:2020nqj,Briceno:2022omu,Briceno:2021xlc}. With a mature formalism at hand, steady progress is being achieved in applications to physically relevant process are being carried out, e.g. weak decay of a kaon to two pions~\cite{RBC:2020kdj}, and form-factor of the meson resonances~\cite{Feng:2014gba,Briceno:2016kkp,Andersen:2018mau,Radhakrishnan:2022ubg, Alexandrou:2018jbt}.

Transitions that involve more than two hadrons constitute the present frontier. Before discussing the progress towards this end, we briefly summarize the status of studies of purely hadronic amplitudes. For the past decade, there has been a strong push towards understanding the non-perturbative relation between the spectrum of three particles in a finite-volume and infinite-volume quantities~\cite{Briceno:2012rv,Polejaeva:2012ut,Hansen:2014eka,Hansen:2015zga,Briceno:2017tce,Hammer:2017uqm,Hammer:2017kms,Mai:2017bge,Briceno:2018mlh,Briceno:2018aml, Blanton:2019igq,Pang:2019dfe,Jackura:2019bmu,Romero-Lopez:2019qrt,
Hansen:2020zhy,Blanton:2020gha,Blanton:2020jnm,Pang:2020pkl,Romero-Lopez:2020rdq,Blanton:2020gmf,Muller:2020vtt,Blanton:2021mih,Muller:2021uur,Blanton:2021eyf,Jackura:2022gib,Garofalo:2022pux,Hansen:2024ffk}. In general, these relations provide a method to constrain a short-distance unphysical quantity, that can then be related to physical scattering amplitudes via non-perturbative integral equations~\cite{Hansen:2015zga}. This has motivated a line of research to solve and understand the analytic properties of the integral equations and their solutions~\cite{Jackura:2018xnx,Briceno:2019muc, Jackura:2020bsk,Sadasivan:2021emk,Dawid:2023jrj,Dawid:2023kxu}, which has resulted in some of the most stringent tests of the aforementioned formalism. Given the confidence in the formalism, there have been several applications in the analysis of actual lattice QCD spectra~\cite{Mai:2018djl,Blanton:2019vdk,Fischer:2020jzp,Alexandru:2020xqf,Brett:2021wyd,Mai:2021nul}, including interactions in higher-partial waves~\cite{Blanton:2021llb,Draper:2023boj}, and one determination of the S-wave scattering amplitude~\cite{Hansen:2020otl}. The formal infrastructure to reliably study integral equations in higher partial waves was recently developed in ref.~\cite{Jackura:2023qtp}.~\footnote{For pedagogical introductions into this rapidly evolving field, we point the reader to Refs.~\cite{Briceno:2017max, Hansen:2019nir}. For more recent summaries of the status of the field, see Refs.~\cite{Hanlon:2024fjd,Romero-Lopez:2022usb}. }

Given this progress, the field has turned its attention towards the possibility of studying three-hadron decays~\cite{Muller:2020wjo,Muller:2022oyw,Hansen:2021ofl,Pang:2023jri}. While no applications have been carried out yet, it is to be expected that the formalism can be used to constrain electromagnetic transitions, such as $\gamma^\star \to3\pi$, as well as CP-violating kaon decays, $K  \to3\pi$. In this context, a thorough investigation of the features of the formalism is timely. This includes consistency checks, which have been proven useful in the context of two-particle processes~\cite{Briceno:2019nns,Briceno:2020xxs}.

In this work, we perform an analytical and numerical validation of the relativistic field-theoretic formalism for three-hadron transitions in infinite- and finite-volume~\cite{Hansen:2021ofl}. For simplicity, we use the formalism for identical particles in the isotropic approximation. In particular, we focus
on two limits. First, we consider the case where the two-particle subchannel contains a bound state (dimer). In this limit, we use the Lehmann–Symanzik–Zimmerman (LSZ) reduction formula to show that below the three-particle threshold, the three-particle transition amplitude recovers the transition amplitude to a two-particle system, composed of the two-body bound state ($b$) and one of the particles ($\varphi$). We prove that the resultant two-body amplitude satisfies Watson's theorem, as expected from S matrix unitarity. Furthermore, we show that 
the finite-volume formalism presented in ref.~\cite{Hansen:2021ofl} reduces to the known Lellouch-L\"uscher formalism for two nondegenerate particles. The second scenario considered is that where the theory contains a three-body bound state (trimer). In this limit, we demonstrate that the finite-volume matrix element is equal to the trimer decay constant up to exponentially suppressed errors in the volume. 
We study both of these limits analytically, and we also show numerical evidence for an illustrative example in the limit of vanishing three-body K matrix.

The rest of this work is organized as follows. In \Cref{sec:IF}, we review the key aspects of the infinite-volume formalism and we consider the two limits discussed above. In \Cref{sec:FV}, we review the necessary finite-volume formalism. 
In \Cref{sec:phib_check}, we analytically derive and provide numerical evidence that the two-body quantization condition and the Lellouch-L\"uscher limit can be recovered from the three-particle formalism in the presence of a two-body bound state. Section \ref{sec:trimer} shows the analogous results in the presence of a trimer, while concluding remarks are given in \Cref{sec:conclu}. In \Cref{app:includingK3}, we generalize the results from \Cref{sec:recoveringQC2,sec:recoveringLL} to non-zero values of the three-body K matrix.

\section{Infinite-volume formalism for identical particles }
\label{sec:IF}  

This section provides a recap of the infinite-volume formalism for the relativistic field-theoretic three-particle scattering~\cite{Hansen:2014eka,Hansen:2015zga} and transition~\cite{Hansen:2021ofl} formalism. 

In what follows, we will strictly consider the limit where the three-particle K matrix, $\Kdf$~\footnote{Note that we have dropped the ``df'' (divergence free) notation from $\Kdf$ originally introduced in ref.~\cite{Hansen:2014eka}, given that K matrices must in general be free from on-shell singularities due to S matrix unitarity.}, is fixed to $0$. Lifting this assumption is expected to be straightforward. In this limit, the three-to-three scattering amplitude for three identical particle scattering can be written as 
\begin{align}
\mathcal{M}_{3}=\mathcal{S}\left\{
\mathcal{D}^{(u,u)}
\right\},
\end{align}
where $\mathcal{D}^{(u,u)}$ satisfies an integral equation, known as the ``ladder equation'' which we define below, including all possible pair-wise interactions. Because this amplitude requires one to define a pair and spectator for the initial and final state, we refer to it as the unsymmetrized amplitude. To obtain the full amplitude, one must sum explicitly over the choice of the spectator. This operation is referred as ``symmetrization'', and it is done by $\mathcal{S}$ in the expression above (see Eq.~(39) of \cite{Hansen:2015zga}). 

In what follows, we explore the consequence of the infinite- and finite-volume formalism where the two-particle subsystem can support a bound state, which we call a dimer ($b$). We will consider the dynamics of this bound state with the spectator, which we will refer generically to $\varphi$. As one might expect, the manifestation of the $\varphi b$ system is most readily available in the unsymmetrized amplitudes. As a result, we will only consider the unsymmetrized $\mathcal{D}^{(u,u)}$. 

Before giving a definition of $\mathcal{D}^{(u,u)}$, we define the kinematics of the system. For simplicity, we will restrict our attention to the amplitudes in their center-of-momentum (CM) frame, where the total three-particle state will carry four-momentum $P^\mu =(E,0)=(\sqrt{s},0)$, where $s$ is the Mandelstam variable. All particles involved are identical spinless bosons with mass $m$. The spectators, which will be on-shell, carry four momenta $k^\mu=(\omega_k,\textbf{k}),$ where $\omega_k=\sqrt{m^2+k^2}$. As a result, the two-particle subsystem has a CM energy ($E_{2k}$) defined by $E_{2k}^2={s_{2k}} =(P-k)^2$.

Furthermore, we will only consider the scenario where the scattering amplitude of the two-particle subsystem, $\mathcal{M}_2$, is completely saturated by the $\ell=0$ partial wave. 
This can be written in terms of the two-body phase-shift, $\delta$, in the standard way, 
\begin{align}
\mathcal{M}_2({k})
=\frac{1}{\rho\cot\delta-i\rho},
\label{eq:M2k}
\end{align}
where $k=|\mathbf{k}|$, $\rho$ is the two-particle phase-space defined for identical particles to be
\begin{align}
    \rho = \frac{q^\star_{2k}}{16\pi\sqrt{s_{2k}}},
    \label{eq:rho2body}
\end{align}
and $q^\star_{2k}$ is the relative momentum of the two systems in its CM frame. This can be written as $q^{\star}_{2k}=\frac{1}{2E_{2k}}\lambda^{1/2}(s_{2k},m^2,m^2)$, where $\lambda$ is the K\"all\'en triangle function
\begin{equation}
\lambda(x,y,z) = x^2+y^2+z^2-2(xy+yz+zx).
\end{equation}
Note, each building block on the right-hand side of \Cref{eq:M2k} is a function of the CM two-particle energy, $\sqrt{s_{2k}} $, but we left this implicit to simplify the notation.

 In what follows, we will make the dependence on the total CM energy implicit, while making the spectator momenta explicit. Labeling the initial/final spectator spatial momentum as $\mathbf{k}/\mathbf{p}$, we can define the ladder equation as
\begin{equation}
\mathcal{D}^{(u,u)}(\mathbf{p},\mathbf{k})=
-\mathcal{M}_2({p})
G(\mathbf{p},\mathbf{k})
\mathcal{M}_2({k})
-\mathcal{M}_2({p})\int\frac{d^3\mathbf{r}}{2\omega_r(2\pi)^3}G(\mathbf{p},\mathbf{r})\mathcal{D}^{(u,u)}(\mathbf{r},\mathbf{k}) \;,
\label{eq:D}
\end{equation}
where $G$ is an exchange propagator that is not a standard one. In particular, in order to regulate the integral appearing in the aforementioned equation, it is necessary for $G$ to depend explicitly on a cutoff function, $H$, 
\begin{align}\label{eq:Ginf}
G (\mathbf{p},\mathbf{k})=  \frac{H(p) H(k)}{(E-\omega_p-\omega_k)^2-(\mathbf{p}+\mathbf{k})^2-m^2}.
\end{align}
This expression follows because we have fixed the angular momentum of the two-particle subsystem to be $\ell=0$. 
There is a large class of cutoff functions that one can consider. A fairly common choice, which we adopt here, is a smooth cutoff function of the form
\begin{equation} \label{eq:cutoff}
    H(k) = J\left(x= \frac{E_{2k}^2}{4m} \right) = \begin{cases}0, & x \leq 0, \\ \exp \left(-\frac{1}{x} \exp \left[-\frac{1}{1-x}\right]\right), & 0<x \leq 1, \\ 1, & 1<x .\end{cases}
\end{equation}
The smoothness of the cutoff is relevant for the finite-volume formalism, but in infinite volume, a hard cutoff also has been used~\cite{Dawid:2023jrj}. In this work, we will only use the smooth form.

To solve the integral equation, it is convenient to write this in terms of a function that has fewer singularities, 
\begin{equation}
\label{eq:ddef}
\mathcal{D}^{(u,u)}(\mathbf{p},\mathbf{k})=
\mathcal{M}_2({p})d(\mathbf{p},\mathbf{k})
\mathcal{M}_2({k}),
\end{equation}
where $d$ satisfies the integral equation,
\begin{equation}
d(\mathbf{p},\mathbf{k})
=-G(\mathbf{p},\mathbf{k})
-\int\frac{d^3\mathbf{r}}{2\omega_r(2\pi)^3}G(\mathbf{p},\mathbf{r})
\mathcal{M}_2({r})
d(\mathbf{r},\mathbf{k}) \;.
\label{eq:d}
\end{equation}

For simplicity, we will further restrict to the case where the orbital angular momenta between the spectator and two-particle system is also fixed to $0$. We can do this by partial-wave projecting $d$~\cite{Jackura:2020bsk}. Labeling the resultant amplitude as $d_S$, it is easy to see that by integrating over $\int d\Omega_k/4\pi$, it satisfies
\begin{equation}
d_S({p},{k})
=-G_S({p},{k})
-
\int_0^\infty\frac{r^2\,dr}{\omega_r(2\pi)^2}
G_S({p},{r})
\mathcal{M}_2({r})
d_S({r},{k}) \;.
\label{eq:dS}
\end{equation}
where 
\begin{equation} \label{eq:Gslogs}
G_S(p, k) 
 =-\frac{H(p)H(k)}{4 p k} \log \left(\frac{z(p, k)+i \epsilon-2 p k}{z(p, k)+i \epsilon+2 p k}\right), 
\end{equation}
with $z(p, k)=\left(\sqrt{s}-\omega_k-\omega_p\right)^2-k^2-p^2-m^2$. The resultant integral equation for $d_S$ can be solved using numerical techniques, as already explored in refs.~\cite{Hansen:2020otl,Jackura:2020bsk,Dawid:2023jrj}.

\subsection{$\mathcal{M}_{\varphi b}$ amplitude}
\label{sec:MphibIV}

In this work, we consider a simple two-model where the two-particle subsystem can have a bound state. The reason for this is that we can explore the fact that in a kinematic region, the physical observables can be equally described in terms of two- and three-body dynamics.

We can do this by parametrizing the two-particle scattering amplitude in terms of the leading order effective range expansion, 
\begin{align}
    q^\star_{2k}\cot\delta = -\frac{1}{a},
\end{align}
where $a$ is the scattering length. If $a>0$, the two-body scattering amplitude has a pole below the threshold corresponding to a bound state with mass
\begin{align} 
    m_b \equiv \sqrt{s_b} =  2\sqrt{m^2-\frac{1}{a^2}}. 
    \label{eq:mbdef}
\end{align}
From this, it is easy to read that the state has a binding momentum of  
\begin{align}
\kappa_{b}=1/a.
\end{align}
The binding momentum will play a key role in \Cref{sec:phib_check}. There, we prove that the finite-volume formalism for the three-particle system can be approximated by the two-body formalism in a kinematic regime, and the difference between these formalisms will be exponentially suppressed by $\kappa_b L$, where $L$ is the spatial extent of the finite-volume system.

It is always true that the residue of the amplitude at the pole can be factorized, 
\begin{align}
\label{eq:M2pole0}
\lim_{s_{2k}\to s_b} \mathcal{M}_2({k})(s_{2k}-s_{b})
=-g^2,
\end{align}
and $g$ can be identified with the coupling of the bound state to the scattering states, i.e. the $b\to 2\varphi $ coupling. For the leading order effective range expansion, the coupling can be found to be
\begin{equation}
g=8\sqrt{2\pi\sqrt{s_b}\kappa_{b}}.
\end{equation}

With this, we can then define an amplitude describing the scattering between the bound state and the spectator. For convenience, we will label this as $\mathcal{M}_{\varphi b}$, where $\varphi$ refers to the spectator, and the amplitude describes elastic $\varphi b \to \varphi b$ scattering. It is a function of $s$, which is suppressed here.

As it was presented in ref.~\cite{Jackura:2020bsk}, in the $\mathcal{K}_{3}\rightarrow 0$ limit, $\mathcal{M}_{\varphi b}$ can be obtained from $\mathcal{D}$ following the LSZ procedure. This is done by amputating the external legs associated with the propagating bound state, and dividing by the corresponding couplings of each state, i.e., 
\begin{equation}
\mathcal{M}_{\varphi b}  =  \lim_{s_{2p},s_{2k}\rightarrow s_b}\,\,
\frac{(s_{2p}-s_b)(s_{2k}-s_b)}{g^2} \mathcal{D}_S^{(u,u)}(p,k),
\end{equation}
where $\mathcal{D}_S$ is the $\mathcal{D}$ amplitude after S-wave projection. 
Using the relations between $\mathcal{D}$ and $d$ in \Cref{eq:ddef}, and the limit of the two-body scattering amplitude at the pole, \Cref{eq:M2pole0}, one finds a simple expression which is more amenable for computational evaluation,
\begin{equation}
\label{eq:Mphib_lim}
\mathcal{M}_{\varphi b} 
= g^2 
\lim_{s_{2p},s_{2k}\,\,
\rightarrow s_b} d_S(p,k),
\end{equation}
where $d_S$ is defined by \Cref{eq:dS}.

Because the resultant amplitude is a two-body scattering amplitude, it must satisfy two-body unitarity, which implies that it can be written in the form, 
\begin{equation}
\label{eq:unitarity}
\mathcal{M}_{\varphi b}   
=\frac{1}{\rho_{\varphi b} \cot\delta_{\varphi b} -i\rho_{\varphi b} },
\end{equation}
where $\delta_{\varphi b}$ is the $\varphi b$ scattering phase shift, $\rho_{\varphi b}$ is its phase-space, 
\begin{align} \label{eq:rhophib}
    \rho_{\varphi b}  = \frac{q_{\varphi b} }{8\pi\sqrt{s}},
\end{align}
and $q_{\varphi b}$ is the relative momentum in the system
\begin{equation}
\label{eq:qphib}
q_{\varphi b} = \frac{1}{2E}\lambda^{1/2}(s,s_b,m^2).
\end{equation}
At times it is more convenient to introduce the $\varphi b$ K matrix, $\mathcal{K}_{\varphi b}$, which is simply related to the phase shift via 
\begin{align} \label{eq:Kphib}
\mathcal{K}_{\varphi b}^{-1} = \rho_{\varphi b} \cot\delta_{\varphi b}.
\end{align}

\subsection{Trimer poles and residues}
\label{sec:trimer_def}

Just like the two-body scattering amplitude can have poles associated with bound state and even resonances, so can the 3-body amplitude $\mathcal{M}_3$. These pole singularities, associated with three-body bound states or resonances, we will generally refer to as \emph{trimers}. As was shown in great detail in refs.~\cite{Dawid:2023jrj,Dawid:2023kxu}, the relativistic scattering amplitudes presented here do generate trimers bound states and/or resonances depending on the value of the two-body scattering length, $a$. The nature and evaluation of these states with $a$ is quite rich and interesting. Here we are just interested in defining the trimer poles, their residues, and ultimately their couplings to external currents. 

Generally, one can have trimers for theories with or without two-body bound states. For simplicity, we will consider classes of theories where there are both two-body bound states and trimers. As a result, trimers appear as poles in both $\mathcal{M}_{\varphi b}$ and the full amplitude in the $s$ complex plane. Here we are just interested in bound state trimers, which lie below the $\varphi b$ threshold on the real axis (in the physical Riemann sheet). If we label the mass of this state as $m_t$, we can define the pole location $(s_t)$ and the binding momentum of the system $\kappa_t$ via
\begin{align}
    m_t \equiv \sqrt{s_t} =  \sqrt{m^2-\kappa_t^2}
    + \sqrt{m_b^2-\kappa_t^2}. 
    \label{eq:kappatdef}
\end{align}

In this work, we will study trimer and couplings via amplitudes involving the $\varphi b$ system. In other words, we will first apply the LSZ procedure to the three-body scattering amplitudes, and then look for poles in $s$. With this, we define the coupling $t\to \varphi b$ coupling ($\gamma_t)$, analogously to as was done in \Cref{eq:M2pole0} for the two-body bound state, 
 \begin{align}
\label{eq:Mphibpole}
\lim_{s \to s_t} (s-s_{t})\mathcal{M}_{\varphi b}
=-\gamma_t^2.
\end{align}

\subsection{$\mathcal{J} \to 3\varphi$ transition amplitude}
\label{sec:Jto3}

Thus far, we have only thought about amplitudes in the absence of external currents. Although what is being considered here are generic properties of scattering amplitudes, one is tempted to refer to the amplitudes above as describing purely ``\emph{hadronic}'' processes, given the motivation of applying this machinery for lattice QCD calculations. With this in mind, we now turn to reactions involving an external probe. Following the analogy with lattice QCD, we can think of this probe as a proxy for perturbative insertions of the electroweak sector. For simplicity, we will assume that the current ($\mathcal{J}$) is a Lorentz scalar, which has the same quantum numbers of the desired three-particle system. Note that we are restricting our attention to scalar bosons projected to $J^P=0^+$, since the resultant amplitude is assured to be a Lorentz scalar. 

With this in mind, here we review the key pieces introduced in ref.~\cite{Hansen:2021ofl} for describing $\mathcal{J} \to 3\varphi$ transitions. Because we exclusively consider the limit where $\mathcal{K}_3$ is zero, the expressions simplify further. Furthermore, we will simplify the notation relative to the previous work. We label the unsymmetrized transition amplitude in the isotropic limit as $\mathcal{T}$, where we have left the superscript $(u)$ and label ``iso'', which normally emphasizes these two points, implicit. 

In this limit, the amplitude can be defined in terms of $\mathcal{D}^{(u,u)}$ or equivalently $d$ up to a multiplicative purely real function $\mathcal{A}$. This function solely depends on $s$. As with the purely hadronic amplitudes, we will leave the dependence on $s$ implicit. With this, the transition amplitude satisfies
\begin{equation} \label{eq:T3varphi}
\mathcal{T}(p)=\mathcal{L} (p)\,  \mathcal{A}
\end{equation}
where 
\begin{align} \label{eq:Luiso}
\mathcal{L}(p)&=\frac{1}{3}- \mathcal{M}_2(p) \widetilde{\rho}(p)
+\int \frac {d^3\mathbf{k}}{2 \omega_k (2 \pi)^3}  i \mathcal{D}^{(u, u)}(\mathbf{p}, \mathbf{k}) 
\,
i \widetilde{\rho}(k) \\
&= \frac{1}{3} 
- \mathcal{M}_2(p)\widetilde{\rho}(p)
- \mathcal{M}_2(p)
\int_0^\infty\frac{k^2\,dk}{\omega_k(2\pi)^2}
d_S(p,k)
\,\mathcal{M}_2(k)
\,\widetilde{\rho}(k) \;,
\end{align}
where we have evaluated the integral over angles analytically and written the remaining integral in terms of $d_S$. We have also introduced a non-standard version of the phase space that depends on the cutoff function,~\footnote{Note, in the literature this non-standard modification of $\rho$ is typically called $\widetilde{\rho}$, because it originally was introduced in a calculation where the principal-value ($\PV$) prescription of integrals being used.} 
\begin{equation}
\label{eq:rho_tilde}
\widetilde{\rho}(k) = -i\rho(k)
H(k)
\;.
\end{equation}

As with $\mathcal{T}$, usually in the literature $\mathcal{L}$ is explicitly labeled with $(u)$ and label ``iso'', which we drop here because we are exclusively interested in the isotropic amplitudes where the outgoing state has not been symmetrized. 

\subsection{$\mathcal{J} \to \varphi b$ transition amplitude}
\label{sec:Jtophib}

Here we now turn our attention to the consequence for the $\mathcal{T}$ in the previously considered scenario where the two-particle subsystem can have a bound state. In this case, the $\mathcal{J} \to 3\varphi $ amplitude will have a pole in the $s_{2p}$ complex plane at $s_b$. 
Following the same LSZ procedure carried out above for relating $\mathcal{D}$ to $\mathcal{M}_{\varphi b}$, it is clear that the residue will be proportional to the $\mathcal{J} \to \varphi b$ amplitude, which we label as $\mathcal{T}_{\varphi b}$,
\begin{equation}
\lim_{s_{2p}\rightarrow s_b}  (s_{2p}-s_b) \mathcal{T}(p) = -g \, \mathcal{T}_{\varphi b}  \;.
\label{eq:T3phitoTphib}
\end{equation}
Because we have fixed the outgoing two-particle subsystem to be at the pole, the $\mathcal{T}_{\varphi b} $ amplitude only depends on $s$, which we leave implicit.

Using the definition
\begin{align}
\begin{split}
\label{eq:TphibT3phi}
\mathcal{T}_{\varphi b} &=
\lim_{s_{2p}\rightarrow s_b}\frac{-(s_{2p}-s_b)}{g}\mathcal{T}(p) \\
&= 
\lim_{s_{2p}\rightarrow s_b}\frac{-(s_{2p}-s_b)}{g}\left[
\frac{1}{3}
- \mathcal{M}_2(p)\widetilde{\rho}(p)
 \right. \\ 
&\left.\qquad 
- \mathcal{M}_2(p)
\int_0^\infty\frac{k^2\,dk}{\omega_k(2\pi)^2}
d_S(p,k)
\,\mathcal{M}_2(k)
\,\widetilde{\rho}(k)\right]\mathcal{A}, 
\end{split}
\end{align}
it follows that
\begin{align}
\begin{split}
\mathcal{T}_{\varphi b} &= \lim_{s_{2p}\rightarrow s_b}-g\left[\widetilde{\rho}(p)+
\int_0^\infty\frac{k^2\,dk}{\omega_k(2\pi)^2}
d_S(p,k)
\,\mathcal{M}_2(k)
\widetilde{\rho}(k) \right]
\mathcal{A} \;,
\\
&= -g\left[\widetilde{\rho}(q_{\varphi b})+
\int_0^\infty\frac{k^2\,dk}{\omega_k(2\pi)^2}
d_S(q_{\varphi b},k)
\,\mathcal{M}_2(k)
\widetilde{\rho}(k) \right]
\mathcal{A} 
\label{eq:infiniteVTphib}\;.
\end{split}
\end{align}
Here we have made use of the behavior of $\mathcal{M}_2$ at its pole, \Cref{eq:M2pole0}, as well as the fact that when the two-particle subsystem is fixed to be at the pole, the spectator momentum coincides with $q_{\varphi b}$ given in \Cref{eq:qphib}. Introducing $\sigma(p)$ as 
\begin{equation}
    \sigma(p) = \widetilde{\rho}(p)+
\int_0^\infty\frac{k^2\,dk}{\omega_k(2\pi)^2}
d_S(p,k)
\,\mathcal{M}_2(k)
\widetilde{\rho}(k)\;,
\label{eq:sigmadefswave}
\end{equation}
the previous result can be abbreviated as
\begin{equation}
    \mathcal{T}_{\varphi b} = -g \, \sigma(q_{\varphi b})
\mathcal{A} \;.
\end{equation}

Given that this is now a transition amplitude coupling to a two-particle state, we know from Watson's theorem that $\mathcal{T}$ must be proportional to $\mathcal{M}_{\varphi b}$ up to an overal real function. In other words, the unitarity singularities of $\mathcal{T}_{\varphi b} $ are completely given by those of $\mathcal{M}_{\varphi b}$. Explicitly, one can show that
\begin{equation}
    \mathcal{T}_{\varphi b} = -g \mathcal{A} ( 1 +i \cM_{\varphi b} \rho_{\varphi b}  ) \mathcal I,
    \label{eq:watsonphib}
\end{equation}
where $\mathcal I$ is a real function of the energy. Since $1 +i \cM_{\varphi b} \rho_{\varphi b}  = \cM_{\varphi b}  \mathcal K^{-1}_{\varphi b} $ and the K matrix is a real function, this equation suffices to demonstrate the proportionality between $\cM_{\varphi b}$ and $\mathcal T_{\varphi b}$.

To show this, we start by noting that $\cM_2(k)$  can be split as 
\begin{align}
    \cM_2  &=  \frac{-g^2}{s_{2k}- s_b+i\epsilon} + \Delta \cM_2,
    \nn\\
    &=  i g^2 \delta(s_{2k}- s_b) + (\Delta \cM_2)',\label{eq:M2pole_IV}
\end{align} 
where $\Delta \cM_2$ and $(\Delta \cM_2)'$ are smooth functions in this kinematic region. Next, consider $\mathcal I^{(n)}(k)$ to be a smooth function, then\footnote{See Appendix A.1 of ref.~\cite{Briceno:2020vgp} for a proof of this relation.} 
\begin{align}
    \begin{split}
         -\int_0^\infty\frac{k^2\,dk}{\omega_k(2\pi)^2}
G_S(q_{\varphi b},k)
\,\mathcal{M}_2(k)
\mathcal I^{(n)}(k)  &= \mathcal I^{(n+1)}(q_{\varphi b}) \\& -G_S (q_{\varphi b},q_{\varphi b}) (i g^2 \rho_{\varphi b})\mathcal I^{(n)}(q_{\varphi b}).
    \end{split}
    \label{eq:poleintegralI}
\end{align}
Here, the first term on the right-hand side, $\mathcal I^{(n+1)}(q_{\varphi b})$, is also a smooth function resulting from a principal-value integration.
Moreover, the second term results from the pole contribution of the integral, which sets all other quantities on-shell. Note that at this stage, the labeling $\mathcal I^{(n)}$ and $\mathcal I^{(n+1)}$ is arbitrary but convenient for the next step.  

In order to reach \Cref{eq:watsonphib}, we start from 
\Cref{eq:infiniteVTphib} and first identify $\mathcal I^{(0)} \equiv \widetilde{\rho}$. Then we insert the integral equation for $d_S$ in \Cref{eq:dS} recursively to all orders. Using the relation in \Cref{eq:poleintegralI} to all orders, one finds infinite sums of the form 
\begin{equation}
    \mathcal I = \sum_n \mathcal I^{(n)},
\end{equation}
where $\mathcal I$ was previously introduced in \Cref{eq:watsonphib}.  Rearranging all terms in the infinite sum, and using $\mathcal M_{\varphi b} = g^2 d_S(q_{\varphi b},q_{\varphi b})$, one arrives to the result in \Cref{eq:watsonphib}.

\subsection{$\mathcal{J} \to t $ coupling}
\label{sec:Jtot}

Finally, we can consider the limit where there is a trimer present in the theory. As previously discussed, we will exclusively consider theories with both two-body states and trimers. The goal here is to define the coupling ($g_t$) of the trimer to the external current.

As in \Cref{sec:trimer_def}, we can then obtain the $\mathcal{J} \to t $ coupling from the residue of the $\mathcal{T}_{\varphi b}$ amplitude at its pole, 
 \begin{align}
\label{eq:Tphibpole}
\lim_{s \to s_t} (s-s_{t})\mathcal{T}_{\varphi b}
=-\gamma_t \, g_t \;.
\end{align}
Note that the residue factorizes, with one piece corresponding to the ``hadronic'' coupling $\gamma_t$ defined in \Cref{eq:Mphibpole}, and the other to the coupling to the current. 

To derive an expression for the decay constant, $g_t$, we make note of the behavior of the full amplitude in the vicinity of the trimer pole~\cite{Hansen:2016ync, Dawid:2023jrj,Dawid:2023kxu}
\begin{equation} \label{eq:Dgammatrimer}
    \mathcal{D}_S^{(u,u)}(p,k)
     \sim-\frac{\Gamma\left(p\right) {\Gamma}(k)}{s-m_t^2} \;,
\end{equation}
where ${\Gamma}(k)$ are related to the wavefunction of the bound states. Given the relationship between $\mathcal{D}$ and $d$, \Cref{eq:ddef}, it follows that $d$ behaves as 
\begin{equation} \label{eq:dStrimer}
    {d}_S(p,k)
    \sim-\frac{1}{\mathcal{M}_2(p)\,\mathcal{M}_2(k)} \frac{\Gamma\left(p\right) {\Gamma}(k)}{s-m_t^2} \;,
\end{equation}

From \Cref{eq:Mphib_lim,eq:Mphibpole}, it is clear that in the limit that $k$ approaches $q_{\varphi b}$ the ratio of $\Gamma$ and $\mathcal{M}_2$ must be smooth, and in particular it must approach
\begin{equation} 
\lim_{s_{2k}\to s_b}\,\,
\frac{{\Gamma}(k)}{\mathcal{M}_2(k)}
=\frac{\gamma_t}{g}.
\end{equation}
This implies that near the timer pole,
\begin{equation}
    {d}_S(q_{\varphi b},k)
    \sim-\frac{\gamma_t}{g\,\mathcal{M}_2(k)} \frac{ {\Gamma}(k)}{s-m_t^2} \;.
\end{equation}

With this, we are finally able to define the trimer decay constant to be 
 \begin{align}
\label{eq:Tphibpole_gt}
\begin{split}
g_t &= -\frac{1}{\gamma_t}\lim_{s \to s_t} (s-s_{t})\mathcal{T}_{\varphi b}
\\
&=
-
\int_0^\infty\frac{k^2\,dk}{\omega_k(2\pi)^2}
{\Gamma}(k)
\widetilde{\rho}(k) 
\mathcal{A} 
.
\end{split}
\end{align}
The key point is that we can define the $\mathcal{J}\to t$ coupling in terms of functions of the ladder equation and one unknown real function, $\mathcal{A}$, which can in principle be constrained from finite-volume matrix elements. Specifically, $g_t/\mathcal{A}$ can be obtained from $T_{\varphi b}/\mathcal{A}$ and $\mathcal{M}_{\varphi b}$ using \Cref{eq:Tphibpole,eq:Mphibpole}.

\section{Recap of the finite-volume formalism}
\label{sec:FV}

In this section, we provide a summary of the finite volume formalism for the processes $\varphi b \rightarrow \varphi b$, $\mathcal{J} \rightarrow \varphi b$, $3\varphi\rightarrow3\varphi$, and $\mathcal{J}\rightarrow3\varphi$, for a system with zero total 3-momentum $\bf{P}=0$. We will only consider systems in cubic volumes, $V=L^3$, with periodic boundary conditions. For simplicity, we will set all angular momenta to 0 and fix $\mathcal{K}_{3} = 0$, although lifting the latter assumption is straightforward. Because we are fixing $\mathcal{K}_{3} = 0$ for every kinematic variable, this is a special case of the isotropic approximation~\cite{Hansen:2014eka}.

A key point of this work is that for energies below the three-body threshold, the finite-volume quantities must be equally well described using the three-body formalism for $3\varphi$ as well as a two-body formalism for the $\varphi b$ system. Given that the two-body formalism is easier, we begin by reviewing this.

\subsection{Finite-volume formalism for the two-body system }
\label{sec:FVphib}

The quantization condition for a two-body state provides a mapping between the set of finite-volume spectra of the system and its scattering amplitude. In general, this relation relates one energy level to an infinite number of partial-wave amplitudes. In practice, at moderately small energies, most partial waves are consistent with zero or unresolvable. As previously mentioned, we only consider the case where the total angular momentum is zero and the individual particles, including the bound state, are spinless. In this case, the quantization condition gives a one-to-one correspondence between the spectrum and the scattering amplitude $\mathcal{M}_{\varphi b}$.

The quantization condition in the kinematic range $(m_b+m)^2<s<(3m)^3$ can be written in terms of the K matrix as~\cite{Luscher:1986pf, Luscher:1990ux, Leskovec:2012gb}, 
\begin{equation} \label{eq:QC2}
(F_{\varphi b}(E,L))^{-1} + \mathcal{K}_{\varphi b}(E) = 0 \;,
\end{equation}
 where the geometric function $F_{\varphi b}$ is given in terms of the L\"uscher-Riemann zeta function, $\mathcal{Z}$,~\footnote{{$\mathcal{Z}$ can be written as $\mathcal{Z}(x)=\left[\sum_{\mathbf{n}} -{\rm PV }\int d^3{n}\right](\mathbf{n}^2-x^2)^{-1}$, where the $\rm PV$ emphasizes that one must take the principal value contribution from the integral. A useful evaluation of this integral is by introducing an exponential regulator. See Appendix B of ref.~\cite{Baroni:2018iau} for an explicit evaluation of the $i\epsilon$-analogue.}} as $F_{\varphi b}(E,L)=\mathcal{Z}(Lq_{\varphi b}/2\pi)/(8\pi^2 LE)$. 

As seen from \Cref{eq:unitarity,eq:Kphib}, the K matrix and $\mathcal{M}_{\varphi b}$ are easily related by unitarity. This leads to another useful representation of the quantization condition in terms of $\mathcal{M}_{\varphi b}$,
\begin{equation} \label{eq:QC2ieps}
(F^{i \epsilon}_{\varphi b}(E,L))^{-1} + \mathcal{M}_{\varphi b}(E) = 0 \;,
\end{equation}
where $F^{i \epsilon}_{\varphi b}$ is defined analogously to $F_{\varphi b}$, except that integrals are performed using the $i\epsilon$ prescription, rather than the principal-value one. The relation between them is straightforward, ${F^{i \epsilon}_{\varphi b} = F_{\varphi b} + i \rho_{\varphi b}}$.

Given the K matrix, one can find the spectrum in a finite volume by looking for the solutions of \Cref{eq:QC2}. Alternatively, given the finite-volume spectrum, one can use \Cref{eq:QC2} to constrain the K matrix, or equivalently the scattering amplitude. The latter is what is actually done in a lattice QCD study.

Next, we review how the $\mathcal{J}\rightarrow \varphi b$ transition amplitude, $\mathcal{T}_{\varphi b}$, can be constrained from finite-volume matrix elements of the current. Given we are only interested in states that are at rest and that couple to 0 total angular momentum, we only need to consider the $A_1^+$ irreducible representation (irrep) of the cubic group, where the $+$ sign labels the parity. If we label the states in this irrep as $\ket{A_1^+,L}$ and normalize them to unity, the transition amplitude can be non-perturbatively related to the finite-volume matrix elements~\cite{Lellouch:2000pv,Briceno:2015csa} as
\begin{equation}
\label{eq:LL2}
|\mathcal{T}_{\varphi b}|^2 = L^3 \frac{|\bra{A_1^+,L}\mathcal{J}\ket{0}|^2}{|1-i\mathcal K_{\varphi b}\rho_{\varphi b}|^2}\left(\frac{\partial F_{\varphi b}(E,L)^{-1}}{\partial E} + \frac{\partial \mathcal K_{\varphi b}(E)}{\partial E}\right) \;,
\end{equation}
where $\rho_{\varphi b}$ is the two-body phase space given in \Cref{eq:rhophib}, and the derivative are evaluated at the corresponding solution of the quantization condition.

\subsection{Finite-volume formalism for the three-body system}
\label{sec:FV3phi}

We now turn our attention to reviewing the finite-volume formalism for the spectrum and matrix elements of three identical scalar particles. As previously stated, we restrict our attention to the isotropic approximation where $\mathcal{K}_{3}=0$ and fix $\bf{P}=0$. 

In this limiting case, the spectrum satisfies~\cite{Hansen:2014eka}
\begin{equation} \label{eq:QC3}
(F^{\text{\iso}}_3(E,L))^{-1} = 0 \;,
\end{equation}
where $F^{\text{\iso}}_3$ can be written as a matrix element of a matrix defined in the spectator-momentum space,
\begin{equation}
F_3^\text{iso}(E,L) = \sum_{ \mathbf{k},\mathbf{p} } \left[F_3(E,L) \right]_{kp}  \equiv \bra{1}F_3(E,L)\ket{1} \;,
 \label{eq:F3iso}
\end{equation}
where the sum runs over all the allowed values of the incoming and outgoing spectator momenta. These satisfy two conditions.
The first is that they must be $\mathbf{k}=2\pi \mathbf{n}/L$, where $\mathbf{n}$ is an integer triplet, as required by the boundary conditions. Second, given the total energy of the system and the value of the momentum, the cutoff function as defined in \Cref{eq:cutoff} must be nonzero, i.e. $H(k)>0$. The state $\ket{1}$ is a vector in the space of momenta with all components equal to 1, while the $F_3$ matrix can be written in terms of a linear combination of other matrices, 
\begin{align} \label{eq:F3matrix}
F_3 &= \frac{1}{ L^3}\left[\frac{\widetilde{F}}{3}-\widetilde{F}\frac{1}{1+2\omega\mathcal{M}_{2L}\widetilde{G}}2\omega\mathcal{M}_{2L}\widetilde{F}\right]\;,
\end{align}
where we suppress the $E$ and $L$ dependence of the quantities above.\footnote{Note that there are several ways of expressing $F_3$ that are algebraically equivalent, for instance, eq. (39) in ref.~\cite{Hansen:2014eka}.}
The diagonal matrix $\widetilde{F}$ is proportional to the zeta function. A convenient representation uses an exponential regulator~\cite{Kim:2005gf} as defined in Appendix B of ref.~\cite{Briceno:2018mlh}
\begin{align}
    \widetilde{F}_{kp}
    &= \delta_{kp}\lim_{\alpha\rightarrow 0}\frac{1}{2\omega_k}\frac{H(k)}{32\pi^3(E-\omega_k)}\frac{2\pi}{L}\left[\frac{1}{L^3}\sum_{\bm{a}}-\text{PV}\int_{\bm{a}}\right]\frac{e^{\alpha(x^2-r^2)}}{x^2-r^2} \;,\\
    &\equiv \delta_{kp}\frac{1}{2\omega_k}
    {F}_{kk'},
    \label{eq:Fkkp}
\end{align}
where $x^2=(q^{*}_{2,k}L/2\pi)^2$, $\bm{a} = 2\pi \bm{n}_a/L$ with $\bm{n}_a$ being vectors of integers, and $r^2=r_{||}^2+r_{\perp}^2$, 
\begin{align}
r_{||} = \frac{n_{a||}-|\bm{n}_k|/2}{\gamma} \;, \quad r_{\perp}=n_{a\perp}\;,
\end{align}
and where $H(k)$ is defined in \Cref{eq:cutoff}. We have introduced a compact notation for the three-dimensional integral, $\int_{\mathbf{a}} \equiv \int d^3\mathbf{a}/(2\pi)^3$, that we adopt going forward.

The $\mathcal{M}_{2L}$ function can be understood as the finite-volume analog of the infinite-volume two-particle scattering amplitude, \Cref{eq:M2k}. The combination $\omega \mathcal{M}_{2L}$ is also  diagonal in momentum space, 
\begin{align}
\left[\omega\, \mathcal{M}_{2L}\right]_{kp} = 
\delta_{kp}\omega_k\left[\rho\cot\delta-i\rho(1-H)+F\right]_{kk}^{-1},
\label{eq:M2L}
\end{align}
where $\delta$ is the two-body scattering phase shift defined in \Cref{eq:M2k}, $\rho$ is the two-body phase space given in \Cref{eq:rho2body}, and the repeated indices on the right-hand side are not being summed over.

 Finally, the $\widetilde{G}$ matrix parametrizes all the one-particle exchanges in a finite volume, and consequently, it is not a diagonal matrix. Assuming the two-particle system is saturated by the $S$-wave angular momentum, its components are defined as
\begin{align}
\widetilde{G}_{kp} &= \frac{H(k)H(p)}{L^3(2\omega_k)(2\omega_{p})((E-\omega_k-\omega_{p})^2-(\mathbf{k}+\mathbf{p})^2-m^2)} \;.\label{eq:Gtilde}
\end{align}

In the three-body sector, it was recently shown~\cite{Hansen:2021ofl} that the $\mathcal{J}\rightarrow 3\varphi$ transition amplitude defined in \Cref{eq:T3varphi} can be related to finite-volume matrix elements $\bra{A_1^+,L}\mathcal{J}\ket{0}$. Given the aforementioned approximations, the relationship reads
\begin{equation} \label{eq:LL3}
|\mathcal{T}(k)|^2 = L^3 |\bra{A_1^+,L}\mathcal{J}\ket{0}|^2|\mathcal{L}(k)|^2 \frac{\partial (F_3^{\text{iso}}(E,L))^{-1}}{\partial E} \;.
\end{equation}
In the above, $\mathcal{L}$ is as defined in \Cref{eq:Luiso}. 
Using \Cref{eq:T3varphi} for the transition amplitude, this relation can be also be written in terms of the infinite-volume quantity $\mathcal{A}$,
\begin{equation} \label{eq:AisoLL}
|\mathcal{A}|^2 = L^3|\bra{A_1^+,L}\mathcal{J}\ket{0}|^2\frac{\partial (F^{\text{iso}}_3(E,L))^{-1}}{\partial E} \;.
\end{equation}
Note that the normalization of these equations differs from eqs.~(2.85) and (2.79) in ref.~\cite{Hansen:2021ofl}, since we are explicitly considering transitions with the vacuum as initial state.

In the case of the trimer, the expectation is that the coupling $g_t$ can be expressed in terms of appropriate finite-volume matrix elements using the standard normalization as
\begin{equation} \label{eq:gtnorm}
g_t = \sqrt{2EL^3}\bra{A_1^+,L}\mathcal{J}\ket{0} \;,
\end{equation}
up to terms exponentially suppressed in $\kappa_tL$, where $\kappa_t$ is the binding momentum of the trimer defined in \Cref{sec:trimer_def}. Combining that with \Cref{eq:AisoLL}, we can obtain $g_t$ from the finite-volume formalism by taking the following limit,
\begin{equation} \label{eq:gtFV}
g_t^2 = \lim_{L\to \infty} \frac{2E\mathcal{A}^2}{\partial (F_3^{\text{iso}}(E,L))^{-1}/\partial E} \;.
\end{equation}

\section{Proving the equivalence for $(m_b+m)^2<s<(3m)^2$}
\label{sec:phib_check}

Here we provide the first check on the finite-volume formalism presented above. We use the same set of assumptions as above, namely, we consider the isotropic case, $\mathcal K_{3} = 0$, and overall zero momentum, $\mathbf P=0$. However, we note that the restriction $\cK_3=0$ can be easily lifted, as done in \Cref{app:includingK3}. Furthermore, we assume that the two-body system has a bound state with mass $m_b$, and consider the consequence of the finite-volume formalism for energies restricted to be above the bound-state/spectator threshold but below the three-particle threshold, i.e. $(m_b+m)^2<s<(3m)^2$. In this kinematic region, both formalisms presented in \Cref{sec:FVphib,sec:FV3phi} must be equivalent.

\subsection{Recovering the two-body quantization condition }
\label{sec:recoveringQC2}

The first goal of this section is to show equivalence between the three-body and two-body quantization conditions given by \Cref{eq:QC3} and \Cref{eq:QC2}, respectively.~\footnote{This was previously shown in ref.~\cite{Briceno:2012rv} for a non-relativistic derivation of the three-body quantization condition.} In particular, we will show for $(m_b+m)^2<s<(3m)^2$ that 
\begin{equation}
    F_3^\text{iso} = \left[ F_3^\text{iso} \right]^\infty + g^2 \sigma (q_{\varphi b}) \frac{1}{\left(F^{i \epsilon}_{\varphi b}\right)^{-1} + \mathcal{M}_{\varphi b}} \sigma (q_{\varphi b}) \;,
    \label{eq:F32p1}
\end{equation}
where $\left[ F_3^\text{iso} \right]^\infty$ is the infinite-volume analog of $F_3^\text{iso}$, explicitly defined as
\begin{align} \label{eq:F3infinity}
\begin{split}
\left[ F_3^\text{iso} \right]^\infty =& \int_{\mathbf{k}} \frac{1}{2\omega_k} \left( \frac{\widetilde{\rho}(k)}{3} -\widetilde{\rho}(k) \mathcal M_2(k) \widetilde{\rho}(k) \right)  \\&-\int_{\mathbf{k},\mathbf{p} } \frac{\widetilde{\rho}(k)}{2\omega_k}  \mathcal M_2(k) d(\mathbf{k}, \mathbf{p})  \mathcal M_2(p)     \frac{\widetilde{\rho}(p)}{2\omega_{p}}  \;,
\end{split}
\end{align}
and $\sigma$ is the same infinite-volume function as in \Cref{eq:sigmadefswave}.   
Note that in the remaining of this section, we suppress $E$ and $L$ dependence in all quantities.
As we will see, \Cref{eq:F32p1} is exact up to exponentially suppressed errors. This is sufficient to argue that in this kinematic region, the spectrum of the theory, which coincides with the poles of $F_3^\text{iso}$, according to \Cref{eq:QC3}, also satisfies the two-body quantization condition, \Cref{eq:QC2}.

For convenience, we can use an expression for $F_3$ that is equivalent to
\Cref{eq:F3matrix}:
\begin{equation}
F_3 = \frac{F}{2\omega L^3}
\left(\frac{1}{3}
-\mathcal{M}_{2L}F
- \mathcal{M}_{2L} d_L 
\frac{\mathcal{M}_{2L}}{2\omega L^3}
  F
\right)  ,
\label{eq:defF3v2}
\end{equation}
where $F = 2 \omega \widetilde{F}$ as in \Cref{eq:Fkkp},
\begin{equation}
    d_L = - {G} 
    -  {G} \frac{\mathcal{M}_{2L}}{2\omega L^3}
 d_L\;,
    \label{eq:dL}
\end{equation}
and ${G}$ is a matrix version of the function $G(\mathbf{p},\mathbf{k})$ shown in \Cref{eq:Ginf}, defined for discrete values of momenta allowed by the boundary conditions. Note, one can also use the $\widetilde{G}$, defined in \Cref{eq:Gtilde} to define a modified version of $d_L$. The advantage of this is that the infinite volume analog of $d_L$ is exactly the $d$ function defined in \Cref{eq:d}. 

The strategy for this derivation will be to separate the volume dependence of  \Cref{eq:defF3v2}, and isolate parts that can lead to divergences. To do this, we take advantage of two key points. First, because $(m_b+m)^2<s<(3m)^2$, the energy running through $\mathcal M_{2L}$ is assured to satisfy $(P-k)^2< (2m)^2$. This allows one to replace $\mathcal M_{2L}$ with $\mathcal{M}_{2}$ up to exponentially suppressed corrections. One can see this from \Cref{eq:M2L} and looking at the behavior of $i\rho H +{F}$ below threshold, which vanishes exponentially quickly~\cite{Hansen:2015zga}. The second key point is that in this same kinematic region, $\mathcal{M}_2$ has a pole associated with the bound state, \Cref{eq:M2pole_IV}. With this, one gets,
\begin{equation}
    \mathcal M_{2L} \simeq \mathcal M_2 =  \frac{-g^2}{s_{2k}- s_b} + \Delta \cM_2 \, ,\label{eq:M2pole}
\end{equation}
where $\Delta \cM_2$ is the smooth function introduced in \Cref{eq:M2pole_IV}. Because $\Delta \cM_2$ has no poles, it will not contribute to power-law finite-volume effects. 

All other functions appearing in the definition of $F_3$ have no poles in this kinematic region. This includes $G$ and $F$. Furthermore, in this kinematic region, we can use the following relation,
\begin{align}
   F_{kp} &= F_{kp} + i{\delta_{kp}} H(k) \rho( k )  
   - i{\delta_{kp}} H(k) \rho( k )  
   \nn\\
   &= F_{kp}^{i\epsilon}    - i{\delta_{kp}} H(k) \rho( k )
   \nn
   \\
   &\simeq
   - i  {\delta_{kp}} H(k) \rho( k ) = \delta_{kp}\widetilde{\rho}(k).
    \label{eq:Ftilde_sub}
\end{align}
Meanwhile, the $G$ matrix can be treated as a matrix that has no singularities.

With this, it is clear that in the kinematic region being considered the only power-law finite-volume effects in the definition of $F_3$, \Cref{eq:defF3v2}, are given by the pole in $\mathcal M_2$. This appears explicitly in the second and third terms in the parenthesis of \Cref{eq:defF3v2}, but it also appears implicitly in the definition of $d_L$, \Cref{eq:dL}. Our task is to isolate such contributions. 

To proceed in a clear fashion, we begin by looking at the leftmost term of $F_{3}^\text{iso}$ in \Cref{eq:defF3v2}. Given the function being summed has no pole singularities, the sum over momenta can be replaced by an integral 
\begin{align}   
\begin{split}
\sum_{\mathbf{k},\mathbf{p}}
\left[\frac{F}{6\omega L^3}\right]_{kp} &\simeq
\frac{1}{L^3}
\sum_{\mathbf{k}}
\frac{\widetilde{\rho}(k)}{6\omega } \\
 &\simeq
\int_{\mathbf{k}}
\frac{\widetilde{\rho}(k)}{6\omega }
 \in
\left[F_{3}^\text{iso}\right]^\infty \;.
\end{split}
\end{align}
The last line denotes that the entire contribution of this first term can be included in the infinite-volume analog of $F_3^{\text{iso}}$ defined in  \Cref{eq:F3infinity}.

Next, we consider the contribution of the second term in \Cref{eq:defF3v2}, 
\begin{align}
-\sum_{\mathbf{k},\mathbf{p}}
\left[\frac{F}{2\omega L^3}
\mathcal{M}_{2L}F\right]_{kp}
&=-\frac{1}{L^3}\sum_{\mathbf{k},\mathbf{k}',\mathbf{p}}
{F}_{kk'}
\frac{1}{2\omega_{{k}'}}
\mathcal{M}_{2L}
F_{k'p}
\nn\\
&\simeq
-\frac{1}{L^3}\sum_{\mathbf{k}}
\widetilde{\rho}(k)
\frac{1}{2\omega_{{k}}}\, 
\mathcal{M}_{2}\, 
\widetilde{\rho}(k).
\label{eq:F3iso2}
\end{align}

To isolate the finite-volume corrections, we will use the simple identity,
\begin{equation}
  \frac{1}{L^3}  \sum_{\mathbf{k}} = 
  \ 
  \int_{\mathbf{k}}
  +
  \left( \frac{1}{L^3}\sum_{\mathbf{k}} - \int_{\mathbf{k}} \right).
  \label{eq:simple}
\end{equation}
Given the approximations already stated and this identity, it is clear that the power-law finite-volume corrections are all in the sum-integral difference of the pole in $\mathcal{M}_2$. In particular, the key identity we will use is the following,
\begin{align} \label{eq:sumM2L}
  \begin{split}
    \sum_{\mathbf k} \frac{1}{2\omega_k L^3} \mathcal M_{2} 
    &=
    \int_{\mathbf{k}} \frac{1}{2\omega_k} \mathcal M_{2} 
    + \left( \frac{1}{L^3}\sum_{\mathbf{k}} - \int_{\mathbf{k}} \right) \frac{1}{2\omega_k} \mathcal M_{2}
    \\
    &=
    \int_{\mathbf{k}} \frac{1}{2\omega_k} \mathcal M_{2} 
    - g^2\, F^{i \epsilon}_{\varphi b}.
    \end{split}    
\end{align}
where we have used that the sum-integral difference of $\mathcal M_{2}$ is related to $F^{i \epsilon}_{\varphi b}$. This can be seen as follows: 
\begin{align}
    \begin{split}
         F^{i \epsilon}_{\varphi b} &= -\frac{1}{g^2}  \left(\frac{1}{L^3}\sum_{\mathbf{k}} - \int_{\mathbf k} \right) \frac{1}{2\omega_k}\cM_2  
         \\
         &\simeq  \left(\frac{1}{L^3}\sum_{\mathbf{k}} - \int_{\mathbf k} \right) \ \frac{1}{2\omega_k(s_{2k} - s_b +i \epsilon)} 
            \\
         &\simeq  \left(\frac{1}{L^3}\sum_{\mathbf{k}} - \int_{\mathbf k} \right) \ \frac{1}{2\omega_k((P-k)^2 - s_b +i \epsilon)} 
    \end{split}
    \label{eq:Fphibeps}
\end{align}
where, as always, we have ignored terms in the integrand that are smooth. Note that after being placed on shell, $F_{\varphi b}^{i\epsilon}$ still has angular dependence, so explicitly is a function of both the on-shell momentum and an angle $\hat{k}$. This dependence can be carried by spherical harmonics when it is written as a matrix in angular momentum space, i.e.,
\begin{equation}\label{eq:F3spherical}
[F^{i\epsilon}_{\varphi b}]_{\ell m,\ell'm'} = \left(\frac{1}{L^3}\sum_{\mathbf{k}} - \int_{\mathbf k} \right) \ \frac{4\pi Y_{\ell m}(\hat{k})Y^*_{\ell' m'}(\hat{k})}{2\omega_k((P-k)^2 - s_b +i \epsilon)} \left(\frac{k}{q_{\varphi b}}\right)^{\ell+\ell'}.
\end{equation}
In \Cref{sec:FV}, we were interested in the $\ell=m=0$ of this matrix, and as a result, we left the angular momentum indices suppressed. Below, we review how the $\ell=0$ can be recovered from the more general results.

With this, we can then revisit \Cref{eq:F3iso2}, 
\begin{align}
-\sum_{\mathbf{k},\mathbf{p}}
\left[\frac{F}{2\omega L^3}
\mathcal{M}_{2L}F\right]_{kp}
&\simeq
-
\int_{\mathbf{k}}
\widetilde{\rho}(k)
\frac{1}{2\omega_{{k}}}\, 
\mathcal{M}_{2}\, 
\widetilde{\rho}(k)
\nn\\
&\hspace{1cm}-\left( \frac{1}{L^3}\sum_{\mathbf{k}} - \int_{\mathbf{k}} \right)
\widetilde{\rho}(k)
\frac{1}{2\omega_{{k}}}\, 
\mathcal{M}_{2}\, 
\widetilde{\rho}(k).
\label{eq:F3iso2v2}
\end{align}
The one subtlety left to address is that the products of $\widetilde{\rho}(k)$ act as effective endcaps multiplying the pole in $\mathcal{M}_2$, which is the only source of power-law finite-volume effects. As a result, we can evaluate these at the ``on-shell'' condition given by the pole, and in doing so we would only be making errors that are exponentially suppressed. Note that at the on-shell condition, the magnitude of $k$ is fixed to be $q_{\varphi b}$, the relative momentum of the $\varphi b$ system, given by \Cref{eq:qphib}. For the system we are considering, $[\tilde{\rho}(q_{\varphi b})]_{\ell m}=\delta_{\ell 0}\delta_{m0}\tilde{\rho}(q_{\varphi b})$, which truncates all the partial waves in \Cref{eq:F3spherical} besides $\ell=m=0$. This allows one to write 
\begin{align}
-\sum_{\mathbf{k},\mathbf{p}}
\left[\frac{F}{2\omega L^3}
\mathcal{M}_{2L}F\right]_{kp}
&\simeq
-\widetilde{\rho}(q_{\varphi b})\cdot
 \left[-g^2 F^{i\epsilon}_{\varphi b}\right]\cdot\, 
\widetilde{\rho}(q_{\varphi b})+... \;,
\label{eq:F3iso2v3}
\end{align}
where the ellipses denote terms that are included in $[F_3^{\text{iso}}]^{\infty}$, and the dots denote a dot product over the angular momentum space.

From the first and second term of $F_3^{\text{iso}}$, we found terms of zeroth order in $F^{i\epsilon}_{\varphi b}$ that can be included in $[F_3^{\text{iso}}]^{\infty}$, and a term of first order in $F^{i\epsilon}_{\varphi b}$ shown in \Cref{eq:F3iso2v3} that contributes to the finite-volume part. The same holds for the third term of $F_3^{\text{iso}}$ as well, except that it contains a series of terms of increasing order in $F^{i\epsilon}_{\varphi b}$ due to the presence of $d_L$. We thus observe a pattern that emerges for $F_3^{\text{iso}}$.  We can systematically evaluate it using \Cref{eq:simple} for each sum, and expanding the series in terms of the number of contributions of $F^{i\epsilon}_{\varphi b}$,
\begin{align}
F_3^{\text{iso}} = \sum_{n=0}^{\infty} F_{3,(n)}^{\text{iso}} \;,
\end{align}
where $F_{3,(n)}^{\text{iso}}$ denotes the term with $n$ factors of $F^{i\epsilon}_{\varphi b}$.

In order to do that, it is useful to introduce endcap functions that are evaluated on the bound-state pole. The on-shell condition generally fixes the magnitude of the momentum that the spectator carries, but it does not fix the angular dependence. This angular dependence can be accounted for using spherical harmonics. The partial-wave projected on-shell endcaps can be written as 
\begin{equation}
     \sigma_{\ell m} (q_{\varphi b}) = \delta_{\ell 0}\,\delta_{m0} \, \widetilde{\rho}(q_{\varphi b}) +  \int_{\mathbf k} d_{\ell m}(q_{\varphi b}, k) \frac{\mathcal M_{2}(k)}{2\omega_{k}} \widetilde \rho({k}),
     \label{eq:sigmadef}
\end{equation}
where $d$ was given in \Cref{eq:d}, and the S-wave projected $\sigma$ was already introduced in \Cref{eq:sigmadefswave}. Note, generally $d$ is not diagonal in the orbital angular momentum~\cite{Jackura:2023qtp}, but in the special case we are considering, with an S-wave two-body bound state, it is. Note, in the limit we have considered, where the $d$ amplitude is completely saturated by the $\ell=0$ partial wave, so will the endcap.

We can then write the term of order $F^{i \epsilon}_{\varphi b}$,
\begin{align}
F_{3,(1)}^\text{iso}
=&- \tilde{\rho}(q_{\varphi b})\cdot\left[
 -g^2
 F^{i \epsilon}_{\varphi b}
 \right]\cdot\tilde{\rho}(q_{\varphi b})
 \nn \\
 &- \tilde{\rho}(q_{\varphi b})\cdot\left[
 -g^2
 F^{i \epsilon}_{\varphi b}
 \right]\cdot\int_{\mathbf{p}} d(q_{\varphi b},p)\frac{\mathcal{M}_2(p)}{2\omega_{p}}\cdot\tilde{\rho}(p)
 \nn \\
 &- \int_{\mathbf{k}} d(k,q_{\varphi b})\frac{\mathcal{M}_2(k)}{2\omega_{k}}\tilde{\rho}(k)\cdot\left[
 -g^2
 F^{i \epsilon}_{\varphi b}
 \right]\cdot\tilde{\rho}(q_{\varphi b})
 \nn \\
 &- \int_{\mathbf{k}} d(k,q_{\varphi b})\frac{\mathcal{M}_2(k)}{2\omega_{k}}\tilde{\rho}(k)\cdot\left[
 -g^2
 F^{i \epsilon}_{\varphi b}
 \right]\cdot\int_{\mathbf{p}} d(q_{\varphi b},p)\frac{\mathcal{M}_2(p)}{2\omega_{p}}\tilde{\rho}(p)
 \nn \\
=&-\sigma (q_{\varphi b}) \cdot 
 \left[
 -g^2
 F^{i \epsilon}_{\varphi b}
 \right]\cdot 
 \sigma (q_{\varphi b}).
\label{eq:F3iso31}
\end{align}

The first term above is exactly the contribution of the second term of $F_3^{\text{iso}}$ shown in \Cref{eq:F3iso2v3}, while the other three terms appear in the expansion of third term of $F_3^{\text{iso}}$. In the last line above, the four contributions have been written compactly in a single term, making use of the endcaps $\sigma(q_{\varphi b})$. Following the same procedure, it is now easy to see the $n^{\text{th}}$ order term,
\begin{align}
F_{3,(n)}^\text{iso}
&
=
-
 \sigma (q_{\varphi b}) \cdot 
 \left[
 -g^2
 F^{i \epsilon}_{\varphi b}
 \right]
  \cdot 
 \left[
 -\mathcal{M}_{\varphi b}
 F^{i \epsilon}_{\varphi b}
 \right]^{n-1}
 \cdot 
 \sigma (q_{\varphi b}),
\label{eq:F3iso3n}
\end{align} 
where we have used the fact that $\mathcal{M}_{\varphi b} = g^2 d$.

Summing all terms to all orders, one gets,
\begin{align}
F_{3}^\text{iso}
\simeq 
\left[F_{3}^\text{iso}\right]^\infty
+
 g^2\sigma (q_{\varphi b}) \cdot 
\left[
\left(F^{i \epsilon}_{\varphi b}\right)^{-1}
+
\mathcal{M}_{\varphi b}
\right]^{-1}\cdot 
 \sigma (q_{\varphi b}).
 \label{eq:F3iso_gen}
\end{align}

With this, we see that the poles of the correlation function satisfy the standard quantization condition for a two-particle system.
In the limit that only the lowest angular momentum contributes, this reduces to the algebraic expression given in \Cref{eq:QC2ieps}, which we rewrite here for convenience 
\begin{align}
\left(F^{i \epsilon}_{\varphi b}(E,L)\right)^{-1}
+
\mathcal{M}_{\varphi b}(E)
=0.
\end{align}

\subsection{Recovering the two-body Lellouch-L\"uscher formalism }
\label{sec:recoveringLL}

Our goal here is to show the equivalence between \Cref{eq:LL3,eq:LL2}. For clarity, we explicitly rewrite these two equations here, including the equality we prove below,
\begin{align} 
\label{eq:LL3_v2}
 |\bra{A_1^+,L}\mathcal{J}\ket{0}|^2
&= \frac{|\mathcal{T}(k)|^2}{L^3\, }\frac{1}{|\mathcal{L}(k)|^2} \left(\frac{\partial (F_3^{\text{iso}})^{-1}}{\partial E}\right)^{-1}
\\
\label{eq:LL2_v2}
&= \frac{|\mathcal{T}_{\varphi b}|^2}{L^3} \frac{|1-i\mathcal K_{\varphi b}\rho_{\varphi b}|^2}{\left(\frac{\partial F^{-1}_{\varphi b}}{\partial E} + \frac{\partial \mathcal K_{\varphi b}}{\partial E}\right) },
\end{align}
where we have assumed the lowest-lying partial wave dominates the $\varphi b$ system, and the $E$ and $L$ dependence of the quantities is not explicitly shown.

To derive this, we start by writing $F_3^\text{iso}$, as given in \Cref{eq:F3iso_gen}, in the vicinity of a finite-volume pole. In particular, we are interested in the derivation of its inverse. Near the pole, the inverse of $F_3^\text{iso}$ vanishes as
\begin{equation}
    (F_3^\text{iso})^{-1} \simeq  \frac{1}{
    g^2 (\sigma(q_{\varphi b}))^2}\left( \frac{1}{F^{i \epsilon}_{\varphi b}} + \mathcal{M}_{\varphi b}\right).
    \label{eq:F32p2}
\end{equation}

Using this identity, we can write its derivative as,
\begin{align}
    \frac{\partial (F_3^\text{iso})^{-1}}{\partial E}\bigg\rvert_{E= E_n} 
    &=  \frac{1}{
    g^2(\sigma(q_{\varphi b}))^2} \frac{\partial}{\partial E} \left( \frac{1}{ F^{i\epsilon}_{\varphi b}} +  \mathcal{M}_{\varphi b}\right)\bigg\rvert_{E= E_n}
    \nn\\
    &=  \frac{-\mathcal{M}_{\varphi b}^2}{
    g^2(\sigma(q_{\varphi b}))^2} \frac{\partial}{\partial E} \left( F^{i\epsilon}_{\varphi b} +  \mathcal{M}^{-1}_{\varphi b}\right)\bigg\rvert_{E= E_n}
    \nn\\
    &=  \frac{|1-i\mathcal K_{\varphi b}\rho_{\varphi b}|^{-2}}{
    g^2|\sigma(q_{\varphi b})|^2} \frac{\partial}{\partial E} \left( F^{-1}_{\varphi b} +  \mathcal{K}_{\varphi b}\right)\bigg\rvert_{E= E_n}
    ,
\end{align}
where we have neglected terms that vanish at the pole, and in the last equality we have used that the phases of $\sigma$ and $\cM_{\varphi b}$ cancel due to Watson's theorem (see \Cref{sec:Jtophib}). 

The other building blocks needed to rewrite the terms on the right-hand side of \Cref{eq:LL3_v2} are the $\mathcal{T}$ and $\mathcal{L}$ amplitudes. Both of these have poles in $s_{2k}$ in the kinematics being considered. The exact behavior at the pole for $\mathcal{T}$ is given by \Cref{eq:T3phitoTphib}. For $\mathcal{L}$, we use \Cref{eq:Luiso} to find 
\begin{equation}
    \lim_{k\rightarrow q_{\varphi b}} (s_{2k} - s_b) \mathcal{L}(k) = \lim_{k\rightarrow q_{\varphi b}}  (s_{2k} - s_b) \left(\frac{1}{3} -\mathcal{M}_2(k) \sigma(k)\right) = g^2 \sigma(q_{\varphi b}).
    \label{eq:LuuforLL}
\end{equation} 

Using the expressions it is now just a matter using simple relations to rewrite the right-hand side of \Cref{eq:LL3_v2}
\begin{align} 
\label{eq:LL3_v3}
 |\bra{A_1^+,L}\mathcal{J}\ket{0}|^2
&= \frac{ |(s_{2k} - s_b) \mathcal{T}(k)|^2}{L^3\, }\frac{1/ |(s_{2k} - s_b)\mathcal{L}(k)|^2}{\frac{\partial (F_3^{\text{iso}})^{-1}}{\partial E} } 
\nn\\
&\simeq 
\frac{ |g\mathcal{T}_{\varphi b}|^2}{L^3\, }
\frac{1/|g^2 \sigma(q_{\varphi b})|^2}
{
\frac{| 1-i\mathcal K_{\varphi b}\rho_{\varphi b}|^{-2}}{
    g^2|\sigma(q_{\varphi b})|^2} \frac{\partial}{\partial E} \left(  F_{\varphi b}^{-1} +  \mathcal{K}_{\varphi b}\right)\bigg\rvert_{E= E_n}
} 
\nn\\
&\simeq 
\frac{ |\mathcal{T}_{\varphi b}|^2}{L^3\, }
\frac{| 1-i\mathcal K_{\varphi b} \rho_{\varphi b}|^{2}}
{
 \frac{\partial}{\partial E} \left(  F_{\varphi b}^{-1} +  \mathcal{K}_{\varphi b}\right)\bigg\rvert_{E= E_n},
} 
\end{align}
which recovers \Cref{eq:LL2_v2}.

\begin{figure}[t]
\centering
\includegraphics[width=\linewidth]{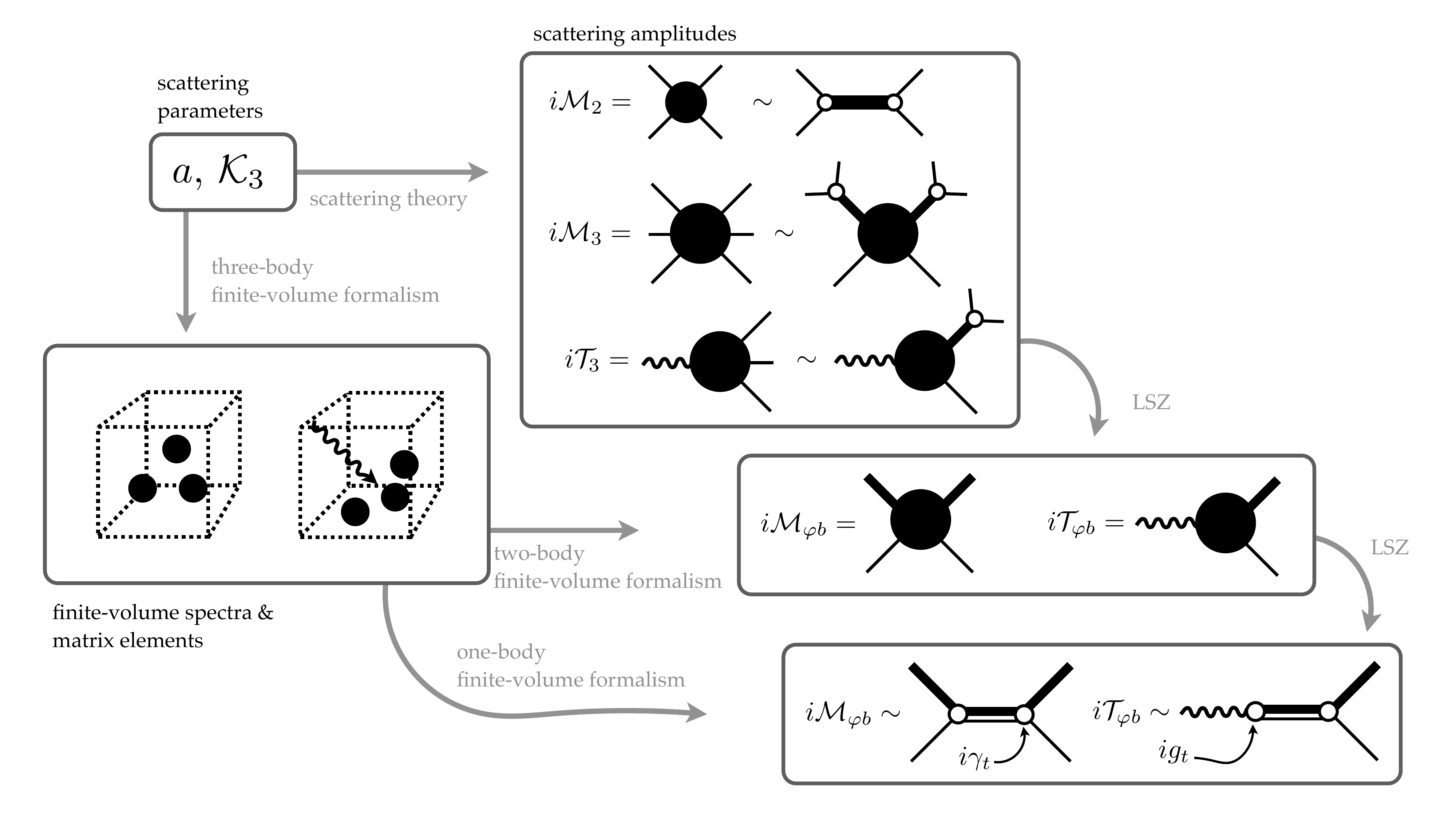}
\caption{A schematic depiction of the workflow used to produce the numerical checks for the amplitudes $\mathcal{M}_{\varphi b}$ and $\mathcal{T}_{\varphi b}$ presented in \Cref{sec:numdimer}, and for the trimer coupling $g_t$ presented in \Cref{sec:numtrimer}. Specifically, $\mathcal{M}_{\varphi b}$, $\mathcal{T}_{\varphi b}$, and $g_t$ can be obtained both using the infinite-volume formalism and LSZ reduction formula as described in \Cref{sec:IF}, and the three- and two-body finite-volume formalisms summarized in \Cref{sec:FV}. This is numerically validated in the results shown in \Cref{fig:FVvsIVformalism,fig:Tcheck,fig:trimerproperties}.
\label{fig:workflow}}
\end{figure}

\subsection{Numerical evidence}
\label{sec:numdimer}

In this section, we provide numerical evidence for the two results that were analytically derived in the previous subsections. In the first part, we explore a simple toy model that supports a two-body bound state. Within this model, we demonstrate two things. First, we show that in the kinematic region below the three-body threshold, the finite-volume spectra and matrix elements are equally well described by the two- and three-body formalism. Second, we produce the same numerical values for the infinite-volume amplitudes, $\mathcal{M}_{\varphi b}$ and $\mathcal{T}_{\varphi b}$, using the finite-volume formalism and the integral equations. These two checks provide empirical evidence of the equivalence found in \Cref{sec:phib_check} and the fact that the finite- and infinite-volume formalisms are self-consistent.

In \Cref{fig:workflow}, we give a graphical depiction of the procedure we used to perform this check, which we proceed to describe. Throughout this numerical demonstration, we set $\Kdf=0$, and fix the two-body phase shift using a single value of the scattering length ${k \cot \delta_{\varphi b} = -1/a}$, with $m a = 1.5$. This leads to a bound state mass of $m_b\simeq 1.49 m$, see \Cref{eq:mbdef}. We focus our attention on energies below the three-body threshold, $E<3m$, where the system should be well described as a three- or two-body system.

With these inputs, we can obtain the three-body amplitudes by solving the integral equations using the now standard techniques described in, for example, refs.~\cite{Jackura:2018xnx, Dawid:2023jrj}. The $\mathcal{M}_{\varphi b}$ and $\mathcal{T}_{\varphi b}$ amplitudes can be obtained using the LSZ procedure, as reviewed in \Cref{sec:IF}.

Having $\mathcal{M}_{\varphi b}$, we can  determine the finite-volume spectrum using the two quantization conditions given in \Cref{eq:QC2,eq:QC3}. Out of convenience, we will introduce two functions of energy and volume, which we will label as QC$_{\varphi b}$ and QC$_{3}$, 
\begin{align}
    {\rm QC}_{\varphi b}(E,L) &=
    F^{-1}_{\varphi b}(E,L) + \mathcal{K}_{\varphi b}(E),
    \\
    {\rm QC}_{3}(E,L) &=(F^{\text{\iso}}_3(E,L))^{-1}.
\end{align}
From the quantization conditions, \Cref{eq:QC2,eq:QC3}, we see that the spectra in a finite volume correspond to the zeroes of these functions. By considering values away from the zeros, we are also able to evaluate the numerical residues needed for the finite-volume matrix elements.

\begin{figure}[t]
\centering
\includegraphics[width=\linewidth]{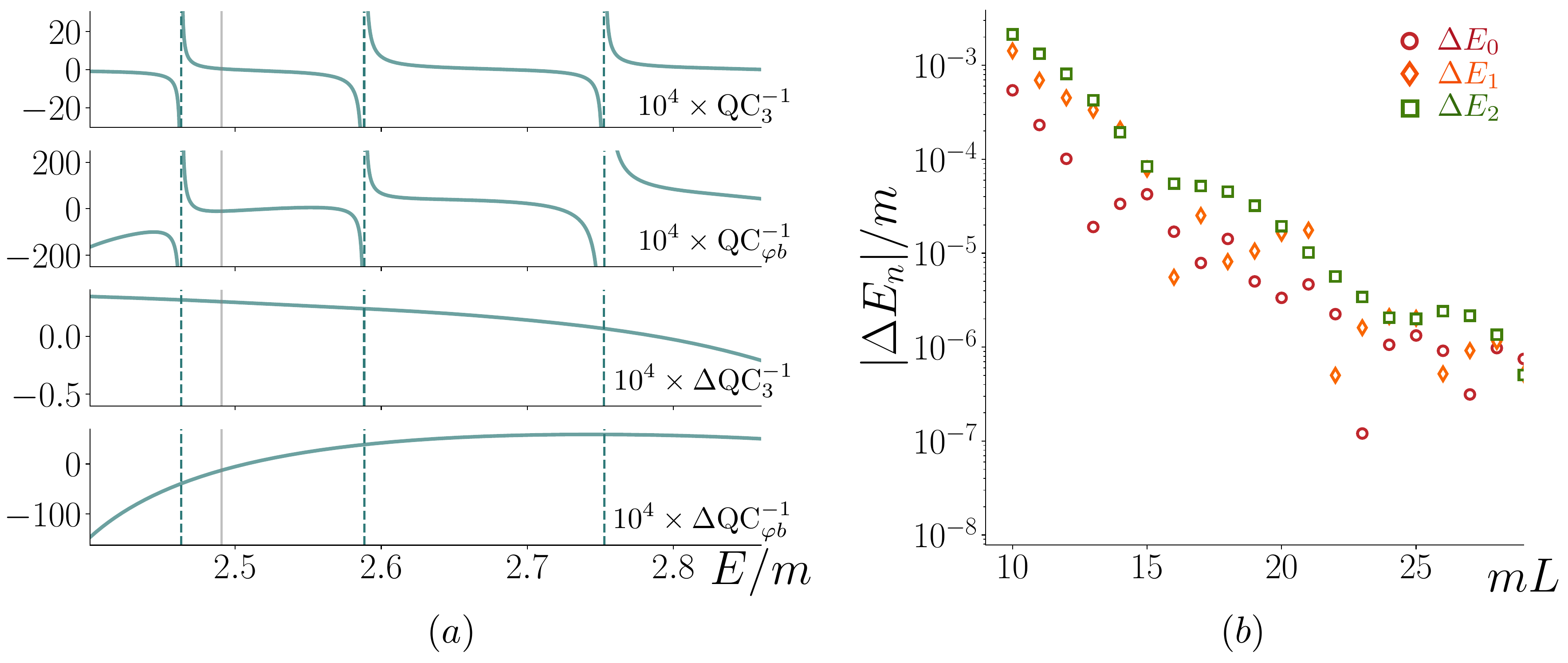}
\caption{$(a)$ Demonstration of the pipeline to obtain energy levels and the corresponding residues of the quantization condition. The top two panels show the inverse of the QCs, while the bottom two show the remainders of the residue determination, defined in \Cref{eq:remainder}, for a fixed volume of $mL=14$. 
$(b)$ The absolute difference between the energies obtained from the two- and three-body quantization conditions as a function of the box size for the three lowest energy levels obtained.
\label{fig:differenceQC2and3}}
\end{figure}

Figure \ref{fig:differenceQC2and3}(a) shows an example of these two functions being evaluated numerically for a volume of $mL=14$. As a stylistic choice, we plot the inverse of these functions, such that the energies are seen as the infinities of ${\rm QC}^{-1}_{3}$ and ${\rm QC}^{-1}_{\varphi b}$. One can immediately see that the locations of the poles of these functions agree by eye. In \Cref{fig:differenceQC2and3}(b) we show the difference between the spectra obtained using these two functions for the three-lowest states for a range of volumes. As can be seen, these spectra agree up to exponentially suppressed errors.

To obtain Lellouch-L\"uscher factors, $\mathcal{R}_{3\varphi,n} $ and $\mathcal{R}_{\varphi b,n} $, given in \Cref{eq:LL2,eq:LL3}, we will need to evaluate the residues of these functions. We do this by evaluating the product of the inverse of the QC functions times $(E - E_{3\varphi,n})$ in the vicinity of a solution, $E \simeq E_{3\varphi,n}$, and interpolating to the desired kinematic point, 
\begin{align}
        \mathcal{R}_{3\varphi,n} &=  
    \lim_{E \to E_{3\varphi,n}}  (E - E_{3\varphi,n})\,{\rm QC}^{-1}_{3}(E,L)\\
    \mathcal{R}_{\varphi b,n} &=  
    \lim_{E \to E_{\varphi b,n}}  (E - E_{\varphi b,n})\,{\rm QC}^{-1}_{\varphi b}(E,L).
    \end{align}
An important subtle point is that although the spectra using ${\rm QC}_{3}$ and ${\rm QC}_{\varphi b}$ are exponentially close to one another, one can not use the spectrum of one in obtaining the residue of the other. These minor differences lead to arbitrarily large systematic errors. Correlatedly, although the pole locations are close, the residues are not.

Finally, to illustrate the quality of the determination of the residues, we introduce two new functions, which are expected to be smooth functions of energy, 
\begin{align}
    \Delta \text{QC}_3 
    &\equiv \text{QC}_3^{-1} - \frac{\mathcal{R}_{3\varphi,n}}{E - E_{3\varphi,n}}, \ 
    \  \\
    \Delta \text{QC}_{\varphi b} 
    &\equiv \text{QC}_{\varphi b}^{-1} - \frac{\mathcal{R}_{\varphi b,n}}{E - E_{\varphi b,n}}. \label{eq:remainder}
\end{align}
In the two lower panels of \Cref{fig:differenceQC2and3}(a), we show the inverse of these functions determined for the same parameters and kinematic region as the QC functions. We see that, as expected, the $\Delta \text{QC}$ functions are indeed smooth in this region. This provides some evidence that the spectra and residues are being obtained with sufficient precision and accuracy.

\begin{figure}[t]
\centering
\includegraphics[width=\linewidth]{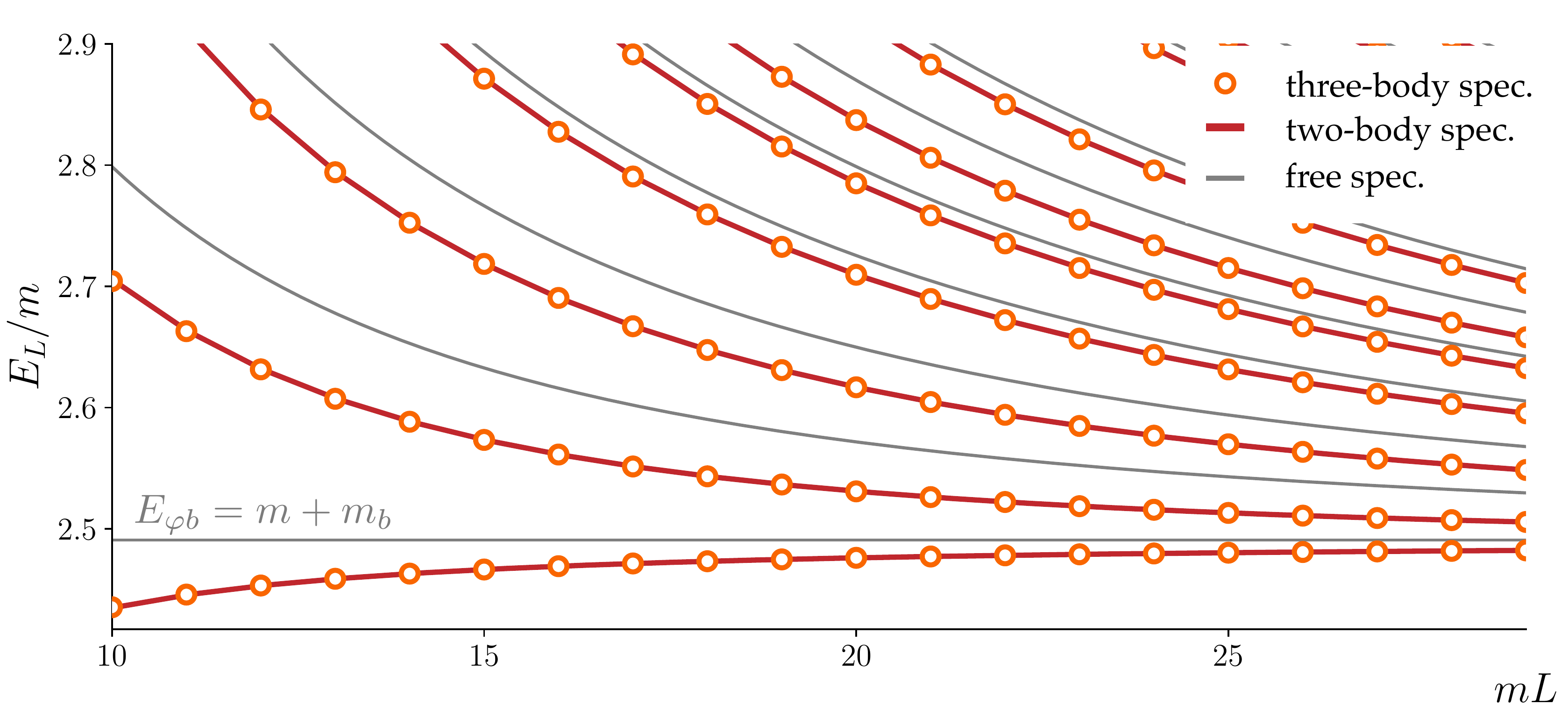}
\caption{Comparison of the energy levels as a function of the box size $L$, obtained from the three-body quantization condition (orange markers) and the two-body one with a particle-dimer phase shift obtained after solving integral equations (red lines). Grey lines indicate the finite-volume energy levels of two particles of mass $m$ and $m_b$ assuming that they do not interact.\label{fig:QC2and3spectrum}}
\end{figure}

Figure \ref{fig:QC2and3spectrum} summarizes the spectra obtained for a large range of volumes using these two methods, which are visibly indistinguishable. This spectrum also shows evidence of a possible shallow three-body bound state below the $\varphi b$ threshold. We will discuss this further below.

\begin{figure}[t]
\centering
\includegraphics[width=\linewidth]{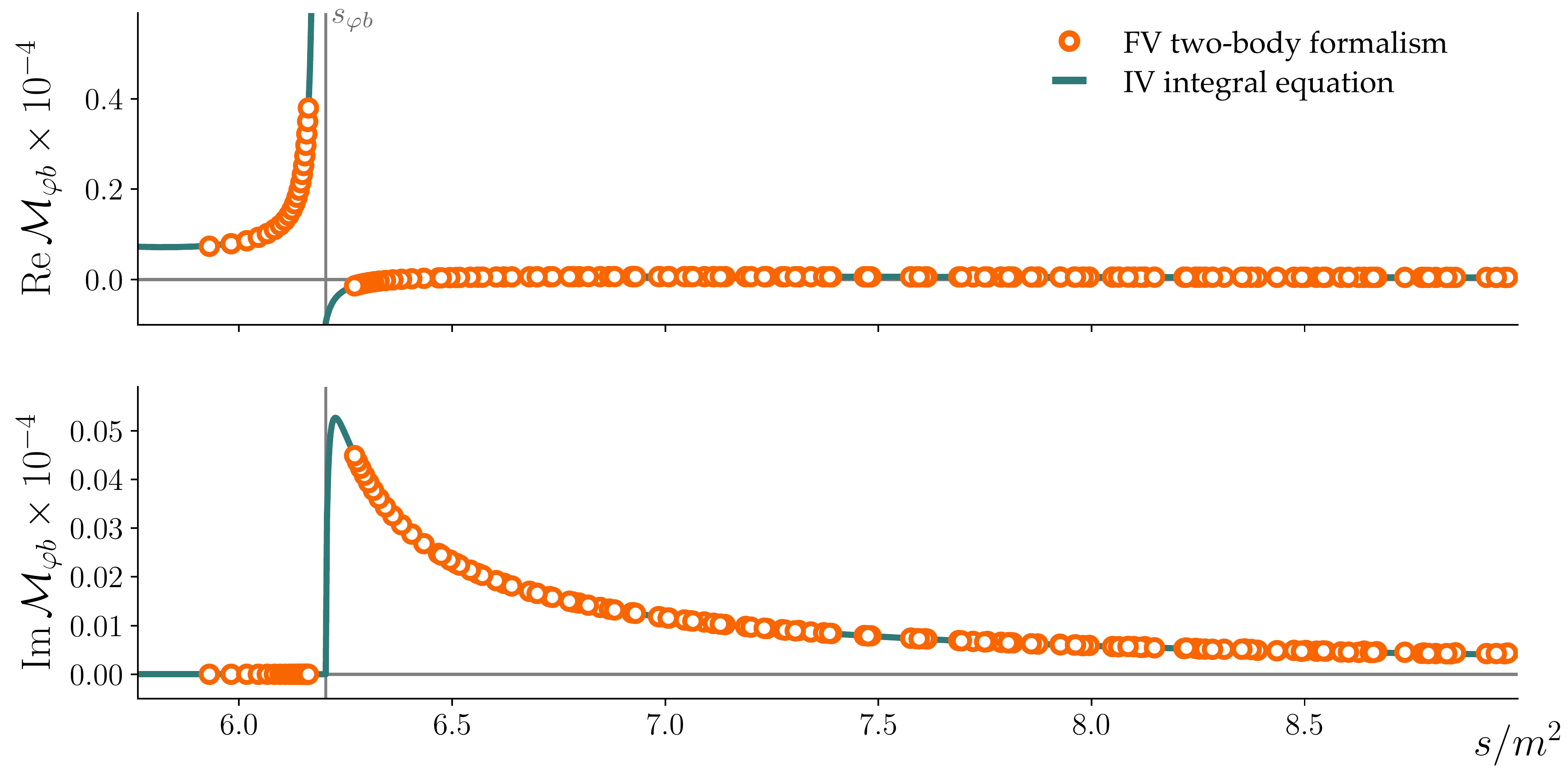}
\caption{Comparison of the real (top) and imaginary (bottom) parts of the particle-dimer scattering amplitude obtained following two methods. The orange dots are obtained by using the two-particle L\"uscher formalism on energies obtained from the three-particle quantization condition. The teal line is obtained by solving the three-body integral equations in combination with the LSZ reduction formula. The vertical grey line labels the particle-dimer threshold, $s_{\varphi b} =(m+m_b)^2$. \label{fig:FVvsIVformalism} }
\end{figure}

Having obtained the spectrum using the three-body quantization condition, we can then input this into the $\varphi b$ quantization condition, \Cref{eq:QC2}, to independently determine the $\cM_{\varphi b}$ amplitude at those energies. In \Cref{fig:FVvsIVformalism}, we show the real and imaginary parts $\cM_{\varphi b}$ as a function of $s=E^2$ obtained using this technique. This procedure only allows for a finite discrete set of determinations of the amplitude, which are shown as orange circles. For comparison, we show the numerical solutions of the integral equations, which can be constrained with arbitrary resolution. As is shown from the figure, one sees perfect agreement between the two methods. This  
 evidence of the self-consistencies of the finite- and infinite-volume formalism was previously observed in refs.~\cite{Jackura:2020bsk,Dawid:2023jrj}.

We are now in place to perform a numerical test of the three-body formalism for decays in the particle-dimer regime. Analogously to the procedure followed for $\cM_{\varphi b}$, we can obtain the $\mathcal J \to \varphi b$ transition amplitude $\mathcal T_{\varphi b}$ following two approaches: directly in infinite volume via integral equations, and using the finite-volume formalism as an auxiliary tool. The infinite-volume method just amounts to feeding the solutions of the integral equation into \Cref{eq:infiniteVTphib}. These approaches are schematically depicted in \Cref{fig:workflow}. Again, this is conceptually straightforward using the techniques presented in ref.~\cite{Jackura:2018xnx, Dawid:2023jrj}.

\begin{figure}[h]
\centering
\includegraphics[width=\linewidth]{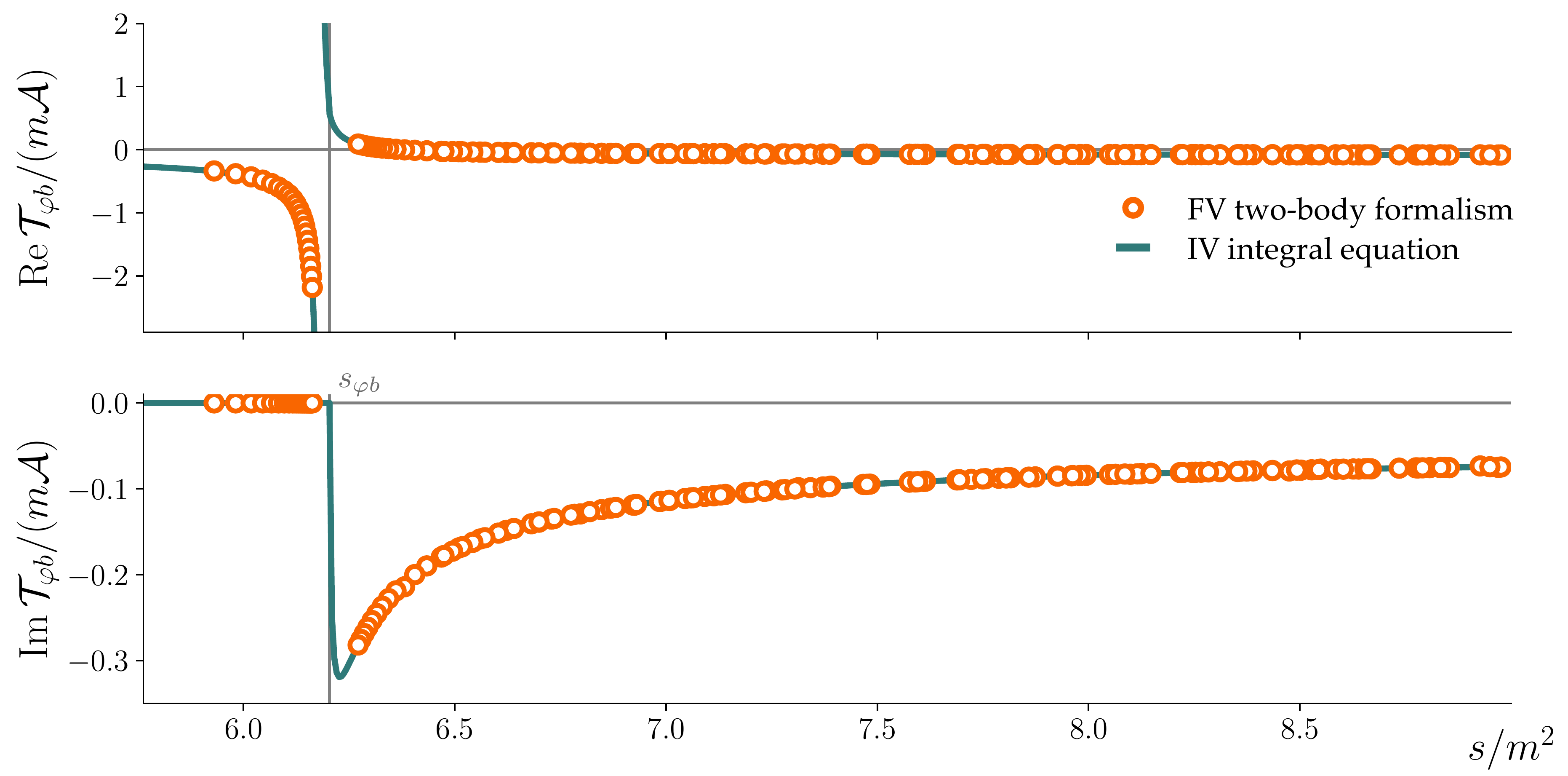}
\caption{Real (top) and imaginary (bottom) parts of the $\mathcal J \to \varphi b$ transition amplitude, $\mathcal T_{\varphi b}$, as a function of $s=E^2$ obtained with two methods. The teal line corresponds to the result obtained by solving the integral equations, while the orange dots are obtained from the finite-volume Lellouch-L\"uscher factors as described in the text. The result is shown in units of the mass $m$ times the arbitrary constant $\mathcal{A}$. The vertical grey line labels the particle-dimer threshold. \label{fig:Tcheck}}
\end{figure}

The finite-volume method follows from the observation that the finite-volume matrix element must satisfy two equalities,
\Cref{eq:LL2,eq:AisoLL}. By equating these two, we can solve for the ratio of $|\mathcal T_{\varphi b }|^2/ |\mathcal A|^2$:
\begin{align}
    \frac{ |\mathcal T_{\varphi b }|^2}{ |\mathcal A|^2} 
    &= 
    \frac{1}{|1-i\mathcal K_{\varphi b}\rho_{\varphi b}|^2}
    \frac{
\frac{\partial ((F_{\varphi b}^{-1} + \mathcal K_{\varphi b})^{-1})}{\partial E}
    }{\frac{\partial (1/F^{\rm iso}_3)}{\partial E}} 
    \\
    &=  \frac{1}{|1-i\mathcal K_{\varphi b} \rho_{\varphi b}|^2} \frac{\mathcal{R}_{3\varphi,n}}{\mathcal{R}_{\varphi b,n}},
\end{align}
where we have written the final expressions in terms of the already determined residua, $\mathcal{R}_{3\varphi,n}$ and $\mathcal{R}_{\varphi b,n}$. This leads to the absolute value of the transition amplitude in units of $\mathcal{A}$. The energy-dependent phase of $\mathcal T_{\varphi b }$ can be inferred using Watson's theorem, i.e., ${\mathcal T_{\varphi b } = |\mathcal T_{\varphi b }| \mathcal M_{\varphi b } / |\mathcal M_{\varphi b }|}$\footnote{There remains a possible overall phase that is not energy dependent and is not fixed by this procedure, either $0$ or $\pi$ to assure that the imaginary part of the amplitude is consistent with unitarity. In order for these two methods to agree, it must be $\pi$. This explains the relative minus sign of the amplitudes shown in \Cref{fig:FVvsIVformalism,fig:Tcheck}.}. 

Figure \ref{fig:Tcheck} also shows the agreement between the finite and infinite-volume determinations of the transition amplitude. As in $\mathcal{M}_{\varphi b}$, we see that $\mathcal{T}_{\varphi b}$ shows evidence of a pole right below threshold, which is consistent with the fact that there is a shallow three-body bound state. This provides perhaps the strongest self-consistency check of the formalism, as well as a numerical validation of Watson's theorem in the three-particle case, which was analytically shown in \Cref{sec:Jtophib}.

\section{Recovering the trimer-coupling}
\label{sec:trimer}
Having analytically continued these results below the three-particle threshold, we can take this continuation further and consider the limit where the three-particle system supports a trimer. 

\subsection{Analytic derivation}

We first comment on the finite-volume effects in the trimer mass. If we consider the trimer as a particle-dimer bound state, the asymptotic volume dependence is known from the two-particle finite-volume formalism
\begin{equation}
    \Delta m_t(L) \propto \frac{1}{L}e^{-\kappa_t L}, \label{eq:trimerFVexp}
\end{equation} 
see e.g. eq.~(42) in Ref.~\cite{Briceno:2019nns}. Since we have already shown that the three-particle formalism reduces to the two-particle one for the particle-dimer system in \Cref{sec:recoveringQC2}, we expect that \Cref{eq:trimerFVexp} describes the leading exponential dependence of the trimer mass.\footnote{Note that \Cref{eq:trimerFVexp} differs from the expression for the volume-dependence of the trimer energy in the unitarity limit~\cite{Meissner:2014dea,Hansen:2016ync}, i.e. $ \Delta m_t(L) \propto \frac{1}{L^{3/2}}e^{-2 \kappa L/\sqrt{3}}$. The apparent discrepancy in the exponent can be understood as a different definition of the binding momentum in refs.~\cite{Meissner:2014dea,Hansen:2016ync}, $m_t = 3m - \kappa^2/m$ (c.f. \Cref{eq:kappatdef}), while the different power of $L$ is a feature of the unitarity limit related to the proximity of the three-particle threshold.} We neglect such exponential effects in the derivation below.

Next, we turn to the formalism for current insertions. Using the definition of the physical decay constant in terms of $\mathcal{A}$ and purely hadronic quantities, we can investigate the finite-volume matrix element in the limit ${E=E_B\ll 3m}$. With $\mathcal{K}_{3}=0$, the finite-volume matrix element is related to $\mathcal{A}$ via  \Cref{eq:AisoLL}, and thus, an expression for $F_3^{\text{iso}}$ in that energy region is needed.

In the trimer limit, \Cref{eq:defF3v2} is dominated by the trimer pole in $d_L$, which is inherited from \Cref{eq:dStrimer}. Thus, we only consider the third term in \Cref{eq:defF3v2}, we can write
\begin{align}
F_3^{\text {iso }}
\simeq 
- 
\sum_{\mathbf{k},\mathbf{p}} \left[\frac{F}{ 2\omega L^3}
 \mathcal M_{2L} d_L  \mathcal M_{2L}
\frac{F}{ 2\omega L^3} \right]_{kp} \simeq \sum_{\mathbf{k},\mathbf{p}} \frac{F_{kk}}{2\omega_k L^3}\frac{\Gamma\left(k\right) \Gamma(p)}{s-m_t^2}
\frac{F_{pp}}{2\omega_{p} L^3},
\end{align}
where we have used \Cref{eq:dStrimer}, and that in the trimer limit $[d_L]_{kp} \simeq  d(\mathbf k, \mathbf p)$. Using the relation for $F_{kk}$ in \Cref{eq:Ftilde_sub}, and replacing sums by integrals, it follows that
\begin{equation}
F_3^{\text {iso }} \simeq
\frac{1}{s-m_t^2}
\left(
\int_{\mathbf{k}} \frac{\widetilde{\rho}(k)}{2 \omega_k} \Gamma(k)\right)^2 \, , \label{eq:F3forgt}
\end{equation}
where, up to exponentially suppressed corrections, sums have been replaced by integrals. Note that the same result can be recovered from \Cref{eq:F32p1}. In particular, below the particle-dimer threshold, $F_{\varphi b}^{i\epsilon}$ vanishes up to exponentially suppressed effects that decay with $\kappa_t$, and thus $F_3^\text{iso}$ is dominated by the second line of \Cref{eq:F3infinity}. However, \Cref{eq:F3forgt} holds even if there is a trimer without a two-particle bound state.

Starting with the definition for $g_t$ in \Cref{eq:Tphibpole_gt}, and using \Cref{eq:F3forgt,eq:AisoLL}, one gets that
\begin{align}
g_t^2    
&=\left(
\int_0^{\infty} \frac{dk\, k^2\,\widetilde{\rho}(k)}{(2\pi)^2\omega_k} \Gamma(k)\right)^2 \mathcal{A}^2
\\
&=
L^3\left\langle A_1^+, L\left|\mathcal{J} \right|0\right\rangle^2 \left(\frac{\partial s}{\partial E}
\right)_{E=E_t}
\\
&=
2EL^3\left\langle A_1^+, L\left|\mathcal{J} \right|0\right\rangle^2 ,
\end{align}
which recovers \Cref{eq:gtnorm}, i.e., the standard relativistic normalization needed to correct for the fact that the finite-volume states have been normalized to 1. In the above, $E=E_t$ is the FV energy corresponding to the trimer state, obtained by solving \Cref{eq:QC3} below the $m_b+m$ energy threshold.

\subsection{Numerical evidence}
\label{sec:numtrimer}

We now turn to a numerical check for the computation of the coupling of a trimer to a current, $g_t$. Following the approach of the previous section, we will use two methods that utilize the finite-volume formalism, as well as the integral equations in the infinite volume. 

First, one can compute $g_t$  from the residue of the pole in $\mathcal T_{\varphi b}$ using \Cref{eq:Tphibpole} after having obtained the trimer pole residue $\gamma_t$ from \Cref{eq:Mphibpole}. Second, $g_t$ can be computed using the QC$_3$ as described in \Cref{eq:gtFV}. This can be rewritten in terms of residue as, 
\begin{equation} \label{eq:gtFV_v2}
g_t^2/\mathcal{A}^2  = \lim_{L\to \infty} 2E_0 \,  \mathcal{R}_{3\varphi,0}\;,
\end{equation}
where $\mathcal{R}_{3\varphi,0}$ is the residue of the lowest finite-volume state. These two approaches are schematically shown in \Cref{fig:workflow}. For \Cref{eq:gtFV_v2} to make sense, the ground state must be below threshold and asymptotic to an energy below threshold as the volume is taken to infinity. From the finite-volume spectra shown in \Cref{fig:QC2and3spectrum}, we see good evidence of there being such a state in our toy model.

\begin{figure}[h]
\centering
\includegraphics[width=\linewidth]{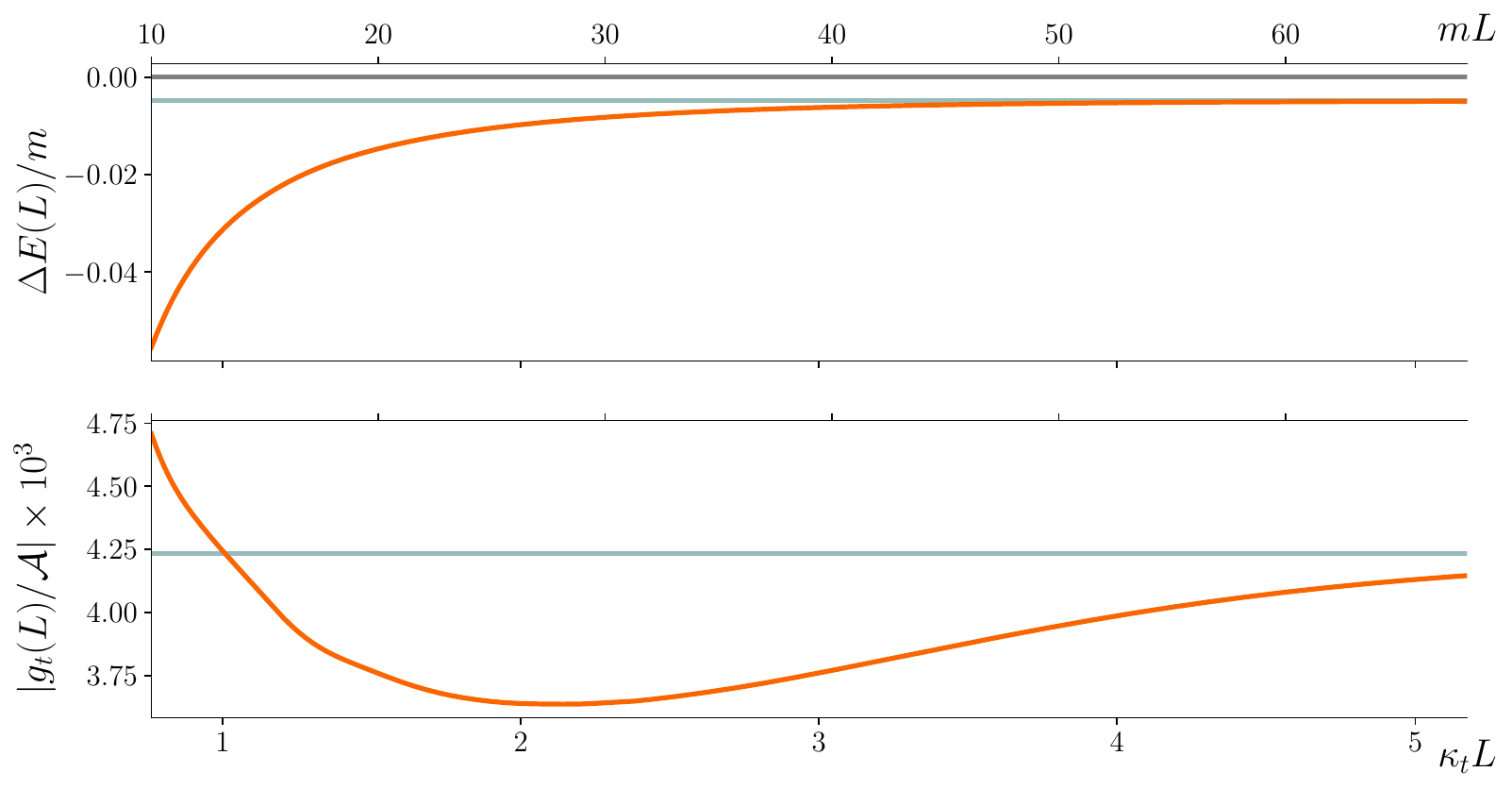}
\caption{Comparisons of properties of the trimer obtained with the finite-volume (orange) and infinite-volume (blue) methods. For the finite-volume method, the orange line shows the energy shift (top) and coupling to the current in units of $\mathcal{A}$ (bottom) of the trimer as a function of the box size in units of the binding momentum of the trimer. The blue lines are the same quantities obtained by solving the integral equations. The top horizontal axis shows the volume in units of $m$, while the bottom horizontal axis shows it units of the binding momentum of the trimer, $\kappa_t$. 
\label{fig:trimerproperties}}
\end{figure}

In \Cref{fig:trimerproperties}, we show the results obtained from determining the energy of this state relative to the $\varphi b$ threshold and the coupling using the finite- and infinite-volume formalisms. For comparison, on the top of the two panels, we show the volume in units of the $m$, and in the bottom axis, we write it in units of the binding momentum of the trimer $\kappa_t$.

As can be seen, both the energy shift and the coupling do approach their infinite-volume values, but the convergence is slow due to the remarkably small binding momentum of the trimer. This is consistent with the expectation that the finite-volume errors should scale with $e^{-\kappa_t L}$ rather than $e^{-m L}$ for a shallow bound state. 
This emphasizes that the finite-volume formalism must be used to treat the volume dependence of such shallow-bounds states. Neglecting to do so would lead to large systematic errors, or it would require a larger order of magnitudes of resources to suppress these exponential effects.

But the key message here is that both the finite- and infinite-volume formalism self consistently recover the same binding energies and matrix elements for three-body bound states when the appropriate limits are taken.

\section{Summary and outlook}
\label{sec:conclu}

In this work, we have performed consistency checks of the infinite- and finite-volume three-body formalism describing three-body scattering and decays derived in ref.~\cite{Hansen:2021ofl,Hansen:2014eka,Hansen:2015zga}. In particular, we have focused in the case in which the two-body subsystem contains a two-body bound state, a dimer. This way, the three-particle formalism below the three-particle threshold describes a particle$+$dimer ($\varphi b$) system.

First, we have shown that in the elastic region of $\varphi b$ scattering, both the three-particle quantization conditions and the formalism connecting finite-volume matrix elements to infinite-volume transition amplitudes reduce to the well-known generalizations of the L\"uscher two-body quantization condition and Lellouch-L\"uscher factor. This was shown first analytically and then numerically in \Cref{sec:phib_check}. The main figures demonstrating this are \Cref{fig:FVvsIVformalism,fig:Tcheck}, where the amplitudes from the integral equations, or via the finite-volume formalism are visibly indistinguishable. Indeed, only small exponentially-suppressed differences are observed.

In \Cref{sec:trimer}, we have also performed numerical and analytical consistency checks in the regime in which there is a three-particle bound state, a trimer, below the $\varphi b$ threshold. In this case, the finite-volume determinations of the trimer mass $m_t$ and coupling to a current $g_t$ must be exponentially close to the same objects obtained from integral equations. This is indeed demonstrated in \Cref{fig:trimerproperties}.

Beyond these consistency checks, this work implements integral equations to compute transition amplitudes for the first time. This is a necessary step towards transitions to three particles from lattice QCD, for which no lattice QCD application are yet available, but are expected in the future.

Phenomenologically relevant three-particle electroweak transitions include $\gamma^* \to 3\pi$ and $K\to3\pi$. These, however, involve three-pion final states at non-maximal isospin. Thus, the generalization to generic three-pion isospin also presented in
refs.~\cite{Hansen:2021ofl,Pang:2023jri} must be used.

As lattice QCD applications keep evolving, it will be necessary to solve integral equation of more complicated systems. While we expect the fundamentals to be simple, this will involve additional challenges, such as multichannel integral equations and higher partial waves. Work in this direction has already started, see refs.~\cite{Jackura:2023qtp}.

Ultimately, the goal is to establish a first-principles pipeline to compute form factors and transition amplitudes involving unstable hadrons, and this work provides a crucial verification for some of the required tools.

\acknowledgments
We thank Max Hansen for useful discussions.
DAP and FRL have been supported in part by the U.S. 
Department of Energy (DOE), Office of Science, Office of Nuclear Physics, under grant Contract Numbers DE-SC000465 (DAP), DE-SC0011090 (FRL) and DE-SC0021006 (FRL). FRL acknowledges support by the Mauricio and Carlota Botton Fellowship. RAB acknowledges the support of the USDOE Early Career award, contract DE-SC0019229. RAB was supported in part by the U.S. Department of Energy, Office of Science, Office of Nuclear
Physics under Awards No. DE-AC02-05CH11231. AWJ acknowledges the support of the USDOE ExoHad Topical Collaboration, contract DE-SC0023598. 


\appendix

\section{Equivalence for non-zero isotropic $\cK_3$}
\label{app:includingK3}

In this appendix, we extend the results of \Cref{sec:recoveringQC2,sec:recoveringLL} to $\mathcal{K}_3 \neq 0$ within the isotropic approximation, i.e., $\mathcal{K}_3$ is simply a function of the total energy.

We first show the equivalence of the two-body and three-body scattering amplitudes in the infinite-volume case. For this, we need the result for the scattering amplitude at non-zero isotropic $\cK_3$ from Ref.~\cite{Hansen:2015zga}:
\begin{equation}
    \cM^{(u,u)}_3(p,k) = \mathcal D_S^{(u,u)}(p,k)  + \frac{\mathcal L(p) \mathcal L(k)}{\left[F^{\rm iso}_3\right]^\infty + \cK_3^{-1} } , 
\end{equation}
where all quantities have been defined in the main text. Applying the LSZ procedure, one can obtain the generalization of \Cref{eq:Mphib_lim} to the non-zero $\cK_3$ case:
\begin{equation}
    \cM_{\varphi b} = g^2 \left( d_S(q_{\varphi b}, q_{\varphi b}) +  \frac{ \sigma(q_{\varphi b})^2}{  \left[F^{\rm iso}_3\right]^{\infty} +  \cK_3^{-1} }  \right),
\end{equation}
where, as in the main text, the energy dependence of the quantities is omitted. 

We now derive the equivalence of the quantization conditions. The three-body quantization condition is~\cite{Hansen:2014eka}
\begin{equation} \label{eq:QC3nonzeroK3}
   F_3^{\rm iso}  = -  \cK_3^{-1}.
\end{equation}

Next, we can use the expression for $F_3^\text{iso}$ derived in section~\ref{sec:recoveringQC2}. In particular,~\Cref{eq:F32p1} describes the relationship between $F_3^\text{iso}$ and $\cM_{\varphi b}$ in the $\Kdf=0$ limit. In this expression, we can substitute $\cM_{\varphi b}$ with its value at $\Kdf=0$, namely $g^2 d_S$, and get
\begin{equation} \label{eq:F3tods}
        F_3^\text{iso} = \left[ F_3^\text{iso} \right]^\infty + g^2 \sigma (q_{\varphi b}) \frac{1}{\left(F^{i \epsilon}_{\varphi b}\right)^{-1} + g^2 d_S(q_{\varphi b},q_{\varphi b})} \sigma (q_{\varphi b}) \;.
\end{equation}

Combining the previous two equations, one can write:
\begin{equation}
    \left(F^{i \epsilon}_{\varphi b}\right)^{-1} + g^2 \left( d_S (q_{\varphi b},q_{\varphi b}) + \sigma(q_{\varphi b})^2 \frac{1}{\left( \left[ F_3^\text{iso} \right]^\infty  + \cK_3^{-1} \right)}
    \right)=  \left(F^{i \epsilon}_{\varphi b}\right)^{-1}  +  \cM_{\varphi b} = 0,
\end{equation}
and thus, the two-body quantization condition is satisfied.

We now prove that three-body formalism for three-body decays is equivalent to the corresponding one for two-body systems below the three-body threshold independently of the value of the isotropic $\cK_3$. To do this, we start with the expression of the finite-volume formalism for decays for non-zero $\cK_3$:
\begin{equation} \label{eq:LLwithK3}
     |\bra{A_1^+,L}\mathcal{J}\ket{0}|^2
= \frac{|\mathcal{T}(k)|^2}{L^3\, }\frac{\left| 1 + \cK_3 \left[ F_3^\text{iso} \right]^\infty \right|^2}{|\mathcal{L}(k)|^2} \left(\frac{\partial (F_3^{\text{iso}})^{-1}}{\partial E} + \frac{\partial \cK_3}{\partial E} \right)^{-1}.
\end{equation}
In order to obtain the relation, it will be useful to express
\begin{equation}
    (F_3^{\text{iso}})^{-1} + \cK_3 = \frac{\left( 1 + \cK_3 \left[ F_3^\text{iso} \right]^\infty  \right)}{ \left[ F_3^\text{iso} \right]^\infty \left\{ \left(F^{i \epsilon}_{\varphi b}\right)^{-1} + g^2 d_S \right\} + g^2 \sigma(q_{\varphi b})^2   } \left( \left(F^{i \epsilon}_{\varphi b}\right)^{-1}  +  \cM_{\varphi b} \right),
\end{equation}
which follows after algebraic manipulations using \Cref{eq:F3tods}. In the vicinity of a finite-volume solution, where the numerator vanishes, the previous equation can be approximated as:
\begin{equation} \label{eq:residuewithK3}
      (F_3^{\text{iso}})^{-1} + \cK_3 \simeq  \frac{1}{g^2 \sigma(q_{\varphi b})^2} \left| 1 + \cK_3 \left[ F_3^\text{iso} \right]^\infty  \right|^2 \left\{ \left(F^{i \epsilon}_{\varphi b}\right)^{-1}  +  \cM_{\varphi b} \right\}, 
\end{equation}
using the two-body quantization condition in the denominator. Note that this also assumes that $\left[ F_3^\text{iso} \right]^\infty$ is real, which is true below the three-body threshold. Plugging in \Cref{eq:residuewithK3} into \Cref{eq:LLwithK3} and using that the derivative is evaluated at the finite-volume solutions, as done in \Cref{sec:recoveringLL}, one can recover \Cref{eq:LL2_v2}.

\bibliographystyle{JHEP} 
\bibliography{biblio}

\end{document}